\documentclass[iop]{emulateapj}
\usepackage{apjfonts}
\usepackage{natbib}
\usepackage{subfigure}
\usepackage{booktabs}

\newcommand{\chandra}{{\it Chandra\/}}
\newcommand{\nustar}{{\it NuSTAR\/}}
\newcommand{\flux}{{erg~cm$^{-2}$~s$^{-1}$}}
\newcommand{\mflux}{{erg~cm$^{-2}$~s$^{-1}$~Hz$^{-1}$}}
\newcommand{\lum}{{erg~s$^{-1}$}}
\newcommand{\mlum}{{erg~s$^{-1}$~Hz$^{-1}$}}

\newcommand{\xray}{\hbox{\it \rm X-ray\/}}

\usepackage{amsmath}

\begin{document}
\title{The Frequency of Intrinsic X-ray Weakness Among Broad Absorption Line Quasars}

\author{
Hezhen~Liu\altaffilmark{1,2,3},
B.~Luo\altaffilmark{1,2,3}, 
W.~N.~Brandt\altaffilmark{4,5,6},
S.~C.~Gallagher\altaffilmark{7},
G.~P.~Garmire\altaffilmark{8}
}

\altaffiltext{1}{School of Astronomy and Space Science, Nanjing University, Nanjing, Jiangsu 210093, China}
\altaffiltext{2}{Key Laboratory of Modern Astronomy and Astrophysics (Nanjing University), Ministry of Education, Nanjing 210093, China}
\altaffiltext{3}{Collaborative Innovation Center of Modern Astronomy and Space Exploration, Nanjing 210093, China}
\altaffiltext{4}{Department of Astronomy and Astrophysics, 525 Davey Lab, The Pennsylvania State University, University Park, PA 16802, USA}
\altaffiltext{5}{Institute for Gravitation and the Cosmos, The Pennsylvania State University, University Park, PA 16802, USA}
\altaffiltext{6}{Department of Physics, 104 Davey Lab, The Pennsylvania State University, University Park, PA 16802, USA}
\altaffiltext{7}{Department of Physics \& Astronomy and Centre for Planetary and Space Exploration, The University of Western
  Ontario, London, ON, N6A 3K7, Canada}
\altaffiltext{8}{Huntingdon Institute for X-ray Astronomy, LLC}

\begin{abstract}
We present combined $\approx 14\textrm{--}37~\rm ks$ \chandra\ observations of
seven $z = 1.6\textrm{--}2.7$ broad absorption line (BAL) quasars selected from
the Large Bright Quasar Survey (LBQS). 
These seven objects are \hbox{high-ionization}
BAL (HiBAL) quasars, and they were undetected in the \chandra\ hard band
(2--8~keV) in previous observations. 
The stacking analyses of previous \chandra\ observations suggested that these seven objects likely contain some candidates for intrinsically \xray\ weak BAL quasars. 
With the new \chandra\ observations, 
six targets are detected. 
We calculate their effective \hbox{power-law} photon indices and \hbox{hard-band} flux weakness,
and find that two objects, \hbox{LBQS $1203+1530$} and \hbox{LBQS $1442-0011$}, show soft/steep spectral shapes ($\Gamma_{\rm eff}= 2.2^{+0.9}_{-0.9}$ and $1.9_{-0.8}^{+0.9}$) and significant \xray\ weakness in the hard band (by factors of $\approx$ 15 and 12). 
We conclude that the two HiBAL quasars are good candidates for intrinsically \xray\ weak BAL quasars. 
The \hbox{mid-infrared-to-UV} spectral energy distributions (SEDs) of the two candidates are consistent with those of typical quasars.
We constrain the fraction of intrinsically \xray\ weak AGNs among HiBAL quasars to be $\approx 7\textrm{--}10\%$ (2/29--3/29), and we estimate it is $\approx 6\textrm{--} 23\%$ (2/35--8/35) among the general BAL quasar population.
Such a fraction is considerably larger than the fraction among non-BAL quasars, and we suggest that intrinsically \xray\ weak quasars are preferentially observed as BAL quasars.
Intrinsically \xray\ weak AGNs likely comprise a small minority of the luminous type~1 AGN population, 
and they should not affect significantly the completeness of these AGNs found in deep \xray\ surveys.
\end{abstract}

\keywords{galaxies: active  -- quasars: absorption lines -- X-rays: general}

\section{Introduction}\label{sec:intro}

 X-ray emission is a characteristic property of active galactic nuclei (AGNs), and it is generally believed to be produced within an \hbox{accretion-disk} corona by \hbox{inverse-Compton} scattering of \hbox{accretion-disk} optical/UV photons \citep[e.g.,][]{Done2010,Gilfanov2014,Fabian2017}.
Previous studies have shown that the UV and \xray\ luminosities of AGNs are related, and this relation is quantified as the \hbox{anti-correlation} between the $2500~{\textup{\AA}}$ monochromatic luminosity ($L_{2500~{\textup{\AA}}}$) and \hbox{X-ray-to-optical} \hbox{power-law} slope ($\alpha_{\rm OX}$)\footnote{$\alpha_{\rm OX}$ is used to compare the optical/UV and \xray\ luminosities and it is defined as 
$\alpha_{\rm OX}=0.3838{\rm log}(f_{\rm 2keV}/f_{\rm 2500~{\textup{\AA}}})$,
where ${\sc f_{\rm 2keV}}$ and $f_{\rm 2500~{\textup{\AA}}}$ are the flux densities 
at \hbox{rest-frame} 2~keV and $2500~{\textup{\AA}}$.} across $\approx 5$ orders of magnitude in UV luminosity \citep[e.g.,][]{Steffen2006,Just2007,Lusso2010}.
This observed relation indicates some underlying physical mechanisms working to balance the emission of accretion disks and coronae. 
Since the observed \xray\ emission could have contributions from AGN jets \citep[e.g.,][]{Miller2011}, and it could also be modified by photoelectric absorption,
studies of the \hbox{$\alpha_{\rm OX}$--$L_{\rm 2500~{\textup{\AA}}}$} relation usually exclude \hbox{radio-loud} AGNs and potentially \xray\ absorbed AGNs, e.g., broad absorption line (BAL) quasars.

Few AGNs are found to be intrinsically \xray\ weak, emitting much less \xray\ radiation than expected from the \hbox{$\alpha_{\rm OX}$--$L_{\rm 2500~{\textup{\AA}}}$} relation. 
For example, \cite{Gibson2008a} systematically investigated the \xray\ and UV properties of optically selected, 
radio-quiet, and non-BAL type 1 quasars. 
Their results showed that the fraction of luminous AGNs that are intrinsically \xray\ weak by a factor of $\approx 10$ is $\lesssim 2\%$.  
Therefore, it has been challenging to identify intrinsically \xray\ weak AGNs. 
One \hbox{well-studied} example is PHL 1811, which is a bright nearby (\hbox{$z=0.192$}) type 1 quasar with a \hbox{{\it B}-band} magnitude of 13.9.
Its observed \xray\ emission is weaker than the expectation from the \hbox{$\alpha_{\rm OX}$--$L_{\rm 2500~{\textup{\AA}}}$} relation by a factor of $\approx 30-100$,
and its soft \xray\ spectrum (with a \hbox{power-law} photon index $\Gamma \approx 2.3$) and \hbox{short-term} \xray\ variability argue against an absorption scenario \cite[e.g,][]{Leighly2007a,Leighly2007b}.
Thus it is believed to be intrinsically \xray\ weak. 
Recently, \xray\ studies of \hbox{less-massive} black-hole (BH) systems ($M_{\rm BH} \approx 10^{4} \textrm{--}10^{6}~M_{\odot}$) have suggested
a few candidates for intrinsically \xray\ weak AGNs \citep[e.g.,][]{Plotkin2016,Simmonds2016}.

The identification of intrinsically \xray\ weak AGNs has considerable importance. 
It challenges the ubiquity of luminous \xray\ emission from AGNs, 
which is central to the utility of \xray\ surveys for finding AGNs throughout the Universe \citep[e.g.,][]{Brandt2015}.
Studies of intrinsically \xray\ weak AGNs might also 
provide insights into the physics of the \xray\ corona.
To obtain the frequency of these rare and extreme AGNs and understand better their nature, it is critical to identify more intrinsically X-ray weak AGNs and to characterize their \xray\ and multiwavelength properties.

There are two criteria for identifying an intrinsically \xray\ weak AGN: 
it must be \xray\ weak (relative to the \hbox{$\alpha_{\rm OX}$--$L_{\rm 2500~{\textup{\AA}}}$} relation) and its \xray\ weakness must be not entirely accounted for by absorption.
BAL quasars are generally observed to be \xray\ weak, with the \xray\ weakness often attributed to absorption due to their often hard \xray\ spectral shapes \cite[e.g.,][]{Gallagher2002a,Gallagher2006,Fan2009,Gibson2009}.
In the accretion-disk wind models for BAL quasars \citep[e.g.,][]{Murray1995,Proga2000}, the BALs are produced in an outflowing wind that is driven by UV radiation pressure. 
In the commonly considered smooth wind scenario \citep{Murray1995}, 
some shielding material is required to protect the wind from overionization by the nuclear ionizing radiation, 
so that the wind can be driven out effectively by radiation pressure.
The observed \xray\ emission could thus be absorbed by the shielding gas. 
In a clumpy wind scenario \citep[e.g.,][]{Baskin2014,Matthews2016},
the wind can be launched successfully without the requirement of 
shielding gas.
However, the clumpy outflows themselves could cause \xray\ absorption.  
Therefore, BAL quasars are often considered to be \xray\ absorbed, and they have usually been excluded in previous searches for intrinsically \xray\ weak AGNs.

On the other hand, if somehow a BAL quasar is intrinsically \xray\ weak, 
the outflowing wind, whether smooth or clumpy, can be launched without fear of overionization, 
leading to the observed BALs and \xray\ weakness.
Indeed, through \nustar\ hard \xray\ (3--24~keV) observations of several significantly \xray\ weak BAL quasars, a few candidates for intrinsically \xray\ weak BAL quasars have been suggested  \citep{Luo2013,Luo2014,Teng2014}; most of these objects have soft effective power-law photon indices ($\Gamma_{\rm eff} \approx 1.8$) in the 3--24~keV band on average (via stacking analysis), disfavoring the absorption scenario.

Motivated by the \nustar\ results, \cite{Luo2013} reinvestigated the \xray\ properties of a \hbox{well-defined} \chandra\ sample of 35 \hbox{high-redshift} ($z \approx 1.5\textrm{--}3$) BAL quasars in \cite{Gallagher2006}, 
which were selected from the Large Bright Quasar Survey (LBQS) sample \citep[e.g.,][]{Hewett1995,Hewett2003}.
At the mean redshift of $z \approx 2$ for these BAL quasars, the $0.5\textrm{--}8$ keV \chandra\ observations sample \hbox{rest-frame} $\approx1.5\textrm{--}24$ keV \hbox{X-rays}, close to the hard \xray\ band observed by \nustar\ for low-redshift targets. 
Stacking analyses of the 12 \hbox{hard-band} (2--8~keV) undetected objects  also revealed a soft effective photon index in the \hbox{rest-frame} $\approx 1.5\textrm{--}24$~keV band \cite[$\Gamma_{\rm eff}\approx 1.6$; e.g., Figure~9 of][]{Luo2013}, 
suggestive of the presence of intrinsically \xray\ 
weak BAL quasars among this \hbox{high-redshift} sample.
Based on these results, a fraction of $\approx 17\textrm{--}40\%$ for intrinsically \xray\ weak AGNs among BAL quasars was estimated, which is apparently much higher than that of intrinsically \xray\ weak AGNs among \hbox{non-BAL} quasars \citep{Gibson2008a}. 

The presence of intrinsically \xray\ weak BAL quasars among the LBQS sample was suggested by the \cite{Luo2013} stacking analyses. 
We could not identify such objects individually, and their fraction among BAL quasars was poorly constrained.
In this paper, we analyze additional \chandra\ observations of seven of the 12 \hbox{hard-band} undetected BAL quasars, including six new \chandra\ Cycle 16 observations and one new \chandra\ archival observation. 
We introduce \chandra\ identification of intrinsic \xray\ weakness and sample selection in Section~\ref{sec:sample}. 
The \xray\ data analyses are described in Section~\ref{sec:data}. 
We present the results and the two good candidates for intrinsically \xray\
 weak BAL quasars in Section~\ref{sec:result}. 
We discuss implications in Section~\ref{sec:discuss},
 and we summarize in Section~\ref{sec:sum}.
 Throughout this paper,
we use J2000 coordinates and a cosmology with
$H_0=67.8$~km~s$^{-1}$~Mpc$^{-1}$, $\Omega_{\rm M}=0.308$,
and $\Omega_{\Lambda}=0.692$ \citep{Ade2016}.

\section{IDENTIFICATION OF INTRINSICALLY X-RAY WEAK AGNs AND SAMPLE SELECTION} \label{sec:sample}
\begin{figure}
\centerline{
\includegraphics[scale=0.42]{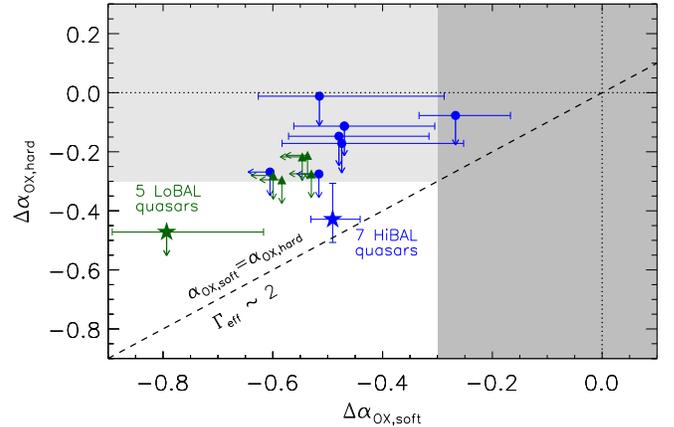}
}
\caption{$\Delta\alpha_{\rm OX,hard}$ versus $\Delta\alpha_{\rm OX,soft}$
for the 12 \hbox{hard-band} undetected LBQS BAL quasars in \cite{Gallagher2006}, 
including seven HiBAL and five LoBAL quasars, 
shown as blue dots and green triangles, respectively.
Upper limits on $\Delta\alpha_{\rm OX,soft}$ ($\Delta\alpha_{\rm OX,hard}$) are shown as arrows when objects are undetected in the soft (hard) \xray\ band.
The stacked source of the HiBAL (LoBAL) quasars is shown as the blue star (green star). 
The dark and light shaded regions show the $\approx 95\%$ ($\approx 2\sigma$) \hbox{confidence-level} uncertainties of $\alpha_{\rm OX,exp}$.
The slanted dashed line indicates
$\alpha_{\rm OX,soft}$ = $\alpha_{\rm OX,hard}$ and $\Gamma_{\rm eff}\approx 2$.
Objects lying outside the shaded regions are significantly \xray\ weak,
and objects lying near the slanted line have soft spectra.
}
\label{fig-aox-old}
\end{figure}

\begin{deluxetable*}{lcccccccccccc}
\tabletypesize{\scriptsize}
\tablewidth{0pt}
\tablecaption{\chandra\ Observation Log and Target Properties
}
\tablehead{
\colhead{Object Name}   &
\colhead{Redshift}   &
\colhead{$B_{\rm J}$}   &
\colhead{Observation}    &
\colhead{Observation}   &
\colhead{New}    &
\colhead{Combined}   &
\colhead{$N_{\rm H,Gal}$}   &
\colhead{BI} &
\colhead{\ion{C}{4} EW} &
\colhead{$v_{\rm max}$} &
\colhead{$R^{\ast}/R_{i}$}   \\
%
\colhead{(LBQS B)}&
\colhead{} &
\colhead{} &
\colhead{Star Date} &
\colhead{ID}  &
\colhead{Exposure (ks)} &
\colhead{Exposure (ks)} &
\colhead{($\rm 10^{20}~cm^{-2}$)}      &
\colhead{(\mbox{\,km\,s$^{-1}$})} &
\colhead{($\textup{\AA}$)} &
\colhead{(\mbox{\,km\,s$^{-1}$})} &
\colhead{}       \\
\colhead{(1)}          &
\colhead{(2)}          &
\colhead{(3)}          &
\colhead{(4)}          &
\colhead{(5)}          &
\colhead{(6)}          &
\colhead{(7)}          &
\colhead{(8)}          &
\colhead{(9)}          &
\colhead{(10)}         &
\colhead{(11)}         &
\colhead{(12)}         
}
\startdata
$ 0021-0213$&$ 2.35$&$  18.68$& 2009 Oct 30&$   8918$&$  29.80$&$  36.55$&$   2.71$&  5179&$  40.20$&$    20138$&            $<0.11/...$\\
$ 1203+1530$&$ 1.63$&$  18.70$& 2016 Apr 22&$  17465$&$  11.95$&$  19.02$&$   2.30$&  1517&$  25.40$&$    11702$&            $.../<0.67$\\
$ 1212+1445$&$ 1.63$&$  17.87$& 2016 Apr 13&$  17466$&$  11.17$&$  15.67$&$   2.94$&  3618&$  38.80$&$    19368$&            $0.01/<0.34$\\
$ 1235+1453$&$ 2.70$&$  18.56$&  2016 May 2&$  17467$&$  11.27$&$  17.91$&$   2.31$&  2657&$  19.30$&$    14414$&            $<0.42/<0.27$\\
$ 1442-0011$&$ 2.23$&$  18.24$& 2016 May 18&$  17468$&$  10.67$&$  14.95$&$   3.39$&  5142&$  39.10$&$    22834$&            $-0.20/<0.51$\\
$ 1443+0141$&$ 2.45$&$  18.20$& 2016 May 12&$  17469$&$  10.08$&$  16.02$&$   3.51$&  7967&$  44.00$&$   >25000$&             $<0.07/<0.77$\\
$ 2201-1834$&$ 1.81$&$  17.81$& 2015 May 23&$  17470$&$   8.60$&$  13.69$&$   2.54$&  1612&$  27.30$&$    19682$&           $<-0.45/...$
\enddata
\tablecomments{
Cols. (1)--(2): object name and redshift. 
Col. (3): $B_{\rm J}$-band magnitude, adopted from \citet{Gallagher2006}.
Cols. (4)--(5): Start date and observation ID of the new \chandra\ archival observation (for \hbox{LBQS $0021-0213$}) or Cycle~16 observation (for the other objects).  
Cols. (6)--(7): exposure time of the new observation and combined exposure time of the previous and new observations.
Col. (8): Galactic neutral hydrogen column density \citep{Dickey1990}.
Cols. (9)--(11): BALnicity Index, \ion{C}{4} absorption equivalent width, and 
maximum velocity of blueshifted \ion{C}{4} absorption, adopted from \citet{Gallagher2006}. 
Col. (12): Logarithm of \hbox{radio-to-optical} flux ratio, adopted from \citet{Gallagher2006}. 
$R^{\ast}$ is the logarithm of the ratio between the flux densities at 5~GHz and $2500~{\textup{\AA}}$ \citep{Stocke1992},
and $R_{i}$ is the logarithm of the ratio between the flux densities at 1.4~GHz and the SDSS $i$ band \citep{Ivezi2002}.
Objects with either $R^{\ast}$ or $R_{i}$ larger than 1 are considered to be radio loud.      
}
\label{tbl-obs}
\end{deluxetable*}

\begin{figure*}
 \centerline{
  \includegraphics[scale=0.42]{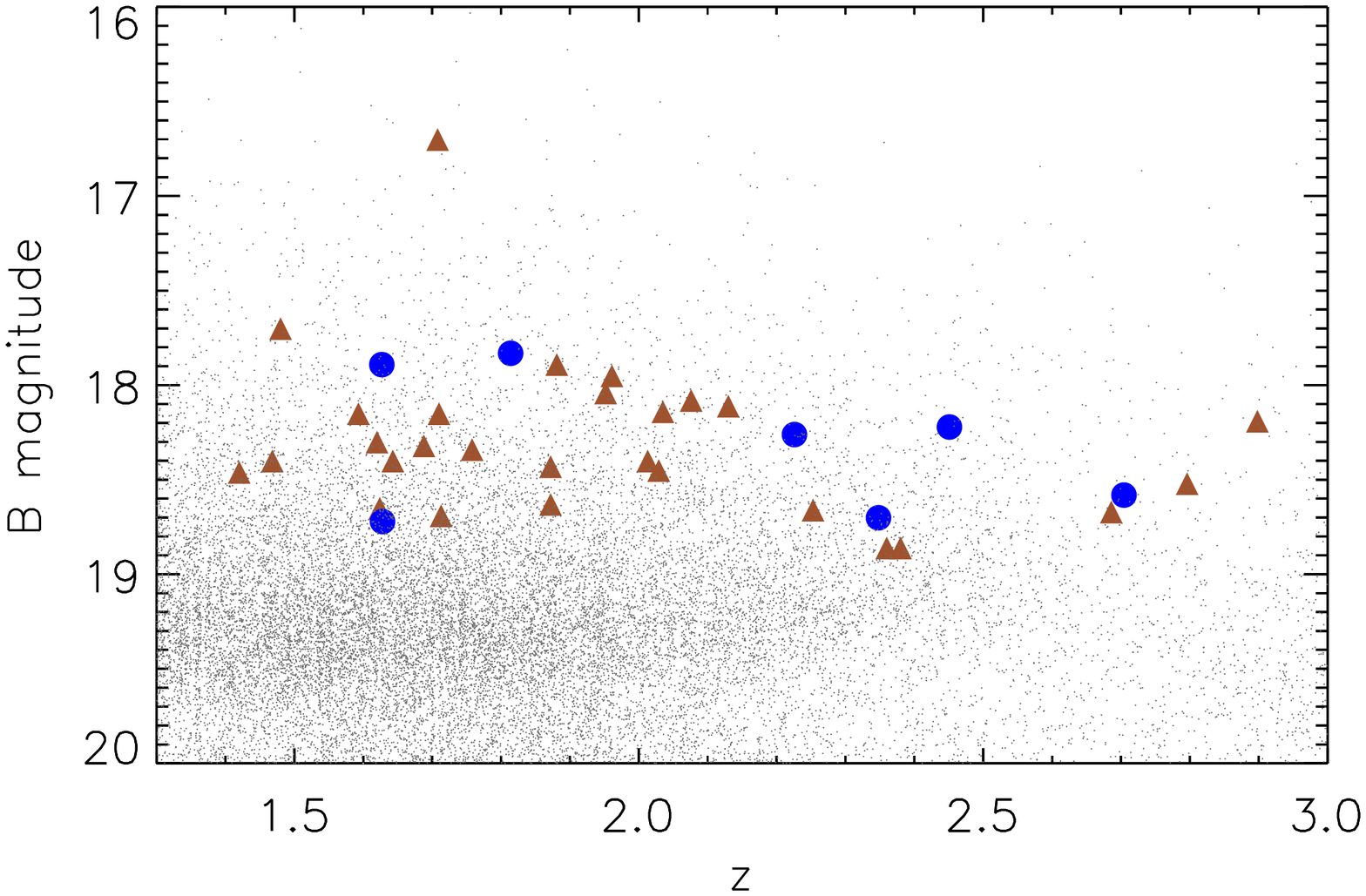}
  \includegraphics[scale=0.42]{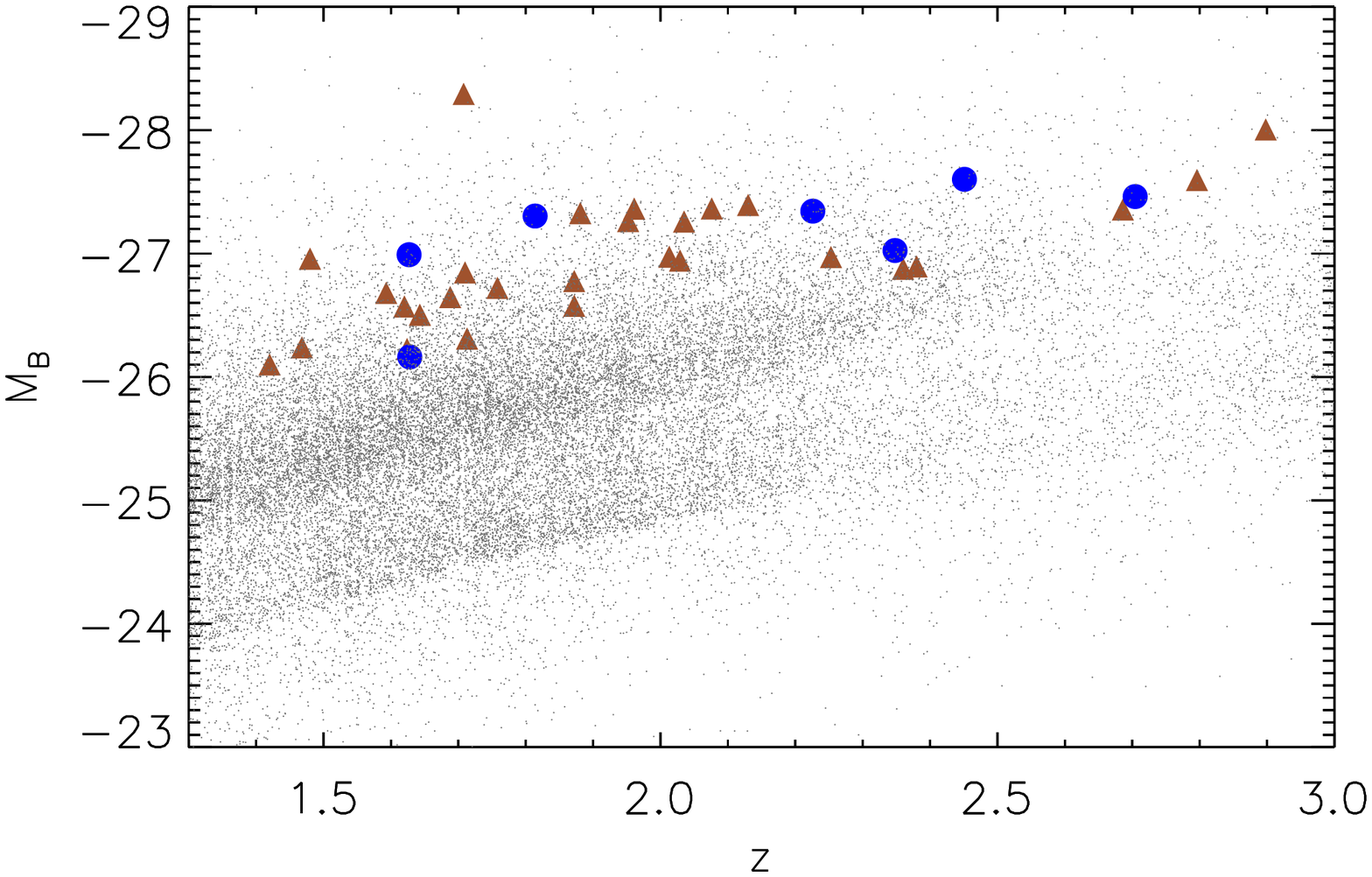}
}
\caption{Redshift versus (a) apparent and (b) absolute $B$-band magnitudes for the 35 BAL quasars in \cite{Gallagher2006}. Our subsample (7 objects) is shown as blue dots and the others are shown as brown triangles.
The underlying small gray dots are objects from the SDSS DR7 quasar catalog \citep{Schneider2010}. The $B$-band magnitudes of the SDSS quasars and LBQS quasars were converted from the $g$-band and $B_{\rm J}$-band magnitudes, respectively, assuming an optical power-law spectrum ($f_{\nu} \propto \nu^{\alpha}$) with a slope of $\alpha= -0.5$ \cite[e.g.,][]{Vanden2001}. 
 \label{fig-lz}} 
\end{figure*}
It has been challenging to identify intrinsically \xray\ weak AGNs,
mainly because it is difficult to determine if the observed \xray\ weakness results from absorption or intrinsic \xray\ weakness.
Spectral fitting of a $\la 8$~keV \xray\ spectrum may not be sufficient to recover the intrinsic \xray\ emission if there is complex absorption present.
For example, if an AGN is obscured 
by a \hbox{Compton-thick} absorber ($N_{\rm H} \ge 1.5 \times 10^{24}~\rm cm^{-2}$), and its $\la 8$ keV \xray\ spectrum is dominated by a possible scattered/reflected soft component,
spectral fitting of the $\la 8$~keV \xray\ spectrum may not necessarily reveal the Compton-thick absorption (see, e.g., \citealt{Comastri2004} for a review).
Therefore, a soft $\la 8$~keV spectrum does not necessarily rule out absorption.
Indeed, the intrinsic \xray\ weakness of PHL 1811 (see Section~\ref{sec:intro}) is supported by both its soft \xray\ spectrum 
and its \hbox{short-term} X-ray variability which argues against a scattering/reflection dominated spectrum.
We have also analyzed a recent 55~ks \nustar\ observation (observation ID 60101004002) of PHL~1811 following the basic procedure in \cite{Luo2014}.
There is a weak detection of this source in the 3--8~keV band, 
but no (or at most a marginal) detection in the 8--24~keV band, 
indicating that there is no strong \hbox{Compton-reflection} component emerging in its hard \xray\ spectrum.

One promising method to identify intrinsically \xray\ weak AGNs is to investigate the hard \xray\ ($\ga 8$~keV) flux weakness and spectral shapes of candidate objects.
In the hard \xray\ band,
due to the combined effects of photoelectric absorption and Compton
scattering, 
a \hbox{Compton-thick} absorber usually results in a flat spectrum with a \hbox{power-law} photon index in a typical range of $\Gamma \approx 0-1$ \cite[e.g.,][]{George1991,Comastri2011,Gandhi2014,Rovilos2014}, 
while for \hbox{radio-quiet} type 1 quasars the mean $\Gamma$
value is $\approx 1.9-2.0$ with an intrinsic dispersion of $\sigma_\Gamma \approx 0.2 \textrm{--}0.3$
\cite[e.g.,][]{Reeves1997,Just2007,Young2009,Mateos2010,Scott2011}.
Based on this consideration, 
\nustar\ observations  
were used to distinguish between the scenarios of \hbox{Compton-thick} absorption and intrinsic \xray\ weakness, 
and a few candidates for intrinsically \xray\ weak BAL quasars were found \citep{Luo2013,Luo2014}.
Stacking analyses indicate that these objects have soft effective photon indices in the $3 \textrm{--}24$~keV band on average, disfavoring the \hbox{Compton-thick} absorption scenario.

At a redshift of $z\approx2$, \chandra\ $0.5\textrm{--}8$~keV observations sample rest-frame energies in the range $\approx 1.5 \textrm{--}24$~keV, similar to the hard \xray\ band probed by \nustar .
Thus, \chandra\ observations of high-redshift targets can also be used to find intrinsically \xray\ weak AGNs.
The hard \xray\ spectral shape can be described by the effective \hbox{power-law} photon index ($\Gamma_{\rm eff}$), derived from the ratio between the \hbox{hard-band} ($2 \textrm{--}8$~keV, \hbox{rest-frame} $\approx 6 \textrm{--}24$~keV) and \hbox{soft-band} ($0.5\textrm{--}2$ keV, \hbox{rest-frame} $\approx 1.5 \textrm{--}6$~keV) count rates.
Moreover, the hard \xray\ flux weakness can be measured via comparing the \hbox{hard-band} derived $\alpha_{\rm OX}$ value ($\alpha_{\rm OX,hard}$, with the rest-frame 2~keV flux density calculated from the observed \hbox{hard-band} flux assuming $\Gamma=2$)
\footnote{Following \cite{Gallagher2006}, we used $\Gamma=2$ instead of the measured $\Gamma_{\rm eff}$ to minimize the effect of potential absorption on the computation of $f_{\rm 2keV}$ and $\alpha_{\rm OX,hard}$. $\Gamma \approx 2$ is the mean photon index for \hbox{radio-quiet} type 1 quasars.} to the expectation ($\alpha_{\rm OX,exp}$) from the $\alpha_{\rm OX}\textrm{--}L_{2500~{\textup{\AA}}}$ relation. 
If a $z\approx2$ AGN is significantly \xray\ weak in the hard band and it also has a soft spectral shape from \chandra\ observations, it is a good candidate for being an intrinsically \xray\ weak AGN.    

\cite{Luo2013} applied the aforementioned method to 
search for intrinsically \xray\ weak AGNs, using the \chandra\ photometric properties of the 35 LBQS BAL quasars in \cite{Gallagher2006}.
These BAL quasars have redshifts of $z \approx 1.5\textrm{--}3$.
There are 23 sources that are detected in the hard band; they have either flat spectral shapes ($\Gamma_{\rm eff}<1$) or $\alpha_{\rm OX,hard}$ values consistent with $\alpha_{\rm OX,exp}$, 
indicating that these 23 sources are likely not intrinsically \xray\ weak AGNs.
Additionally, there are 12 \hbox{hard-band} undetected sources,  
including seven high-ionization BAL (HiBAL) quasars and five \hbox{low-ionization} (LoBAL) quasars.\footnote{HiBAL quasars are objects showing only \hbox{high-ionization} BALs in UV spectra, and LoBAL quasars are objects showing strong \ion{Mg}{2} or \ion{Al}{3} BALs in UV spectra.
Among the 35 BAL quasars in \cite{Gallagher2006}, 
there are 24 HiBAL quasars, six LoBAL quasars, 
and five quasars with unknown BAL types, 
for which the available UV spectra do not cover the \ion{Mg}{2} region. 
HiBAL quasars are the majority population ($\approx 85\%$) of BAL quasars 
 \cite[e.g.,][]{Weymann1991,Sprayberry1992}, and thus 
 we consider the two objects with unknown BAL types among the 12 \hbox{hard-band} undetected objects as HiBAL quasars.}
Figure~\ref{fig-aox-old} shows a comparison of the \hbox{soft-band} flux weakness ($\Delta\alpha_{\rm OX,soft}=\alpha_{\rm OX,soft}-\alpha_{\rm OX,exp}$) to the \hbox{hard-band} flux weakness ($\Delta\alpha_{\rm OX,hard}=\alpha_{\rm OX,hard}-\alpha_{\rm OX,exp}$)  
for the 12 objects, where the \hbox{soft-band} derived $\alpha_{\rm OX,soft}$ is calculated using the \hbox{soft-band} flux and the measured $\Gamma_{\rm eff}$ value.
The shaded regions ($\Delta\alpha_{\rm OX}=0.3$) show the $\approx 95\%$ ($\approx 2\sigma$) uncertainties associated with 
$\alpha_{\rm OX,exp}$ \citep[from Table 5 of][]{Steffen2006}.
Since $\alpha_{\rm OX,soft}$ was computed using the measured $\Gamma_{\rm eff}$ value, and $\alpha_{\rm OX,hard}$ was calculated by assuming $\Gamma=2.0$, 
consistent $\Delta\alpha_{\rm OX,soft}$ and $\Delta\alpha_{\rm OX,hard}$ values indicate $\Gamma_{\rm eff}\approx 2$ (a steep/soft spectral shape; slanted line in Figure~\ref{fig-aox-old}). 
The 12 \hbox{hard-band} undetected objects only have upper limits on $\Delta\alpha_{\rm OX,hard}$, and their \hbox{hard-band} flux weakness and effective photon indices
cannot be constrained individually.
Stacking analyses of the 12 sources were used to constrain their average \xray\ properties.
Stacked flux from the seven HiBAL quasars is significantly detected in both the soft and hard bands,
while stacked flux from the five LoBAL quasars is only detected in the soft band.
In Figure~\ref{fig-aox-old}, the data points for the stacked sources are shown as stars.
The stacked source of the seven HiBAL quasars lies outside the shaded regions (negative $\Delta\alpha_{\rm OX,soft}$ and $\Delta\alpha_{\rm OX,hard}$) and close to the slanted line, 
showing substantial \hbox{hard-band} flux weakness and a soft spectral shape with $\Gamma_{\rm eff}=1.8^{+0.5}_{-0.5}$. 
Thus, this stacked source is likely intrinsically \xray\ weak, which suggests the presence of intrinsically \xray\ weak AGNs among the seven HiBAL quasars.
For the stacked source of the five LoBAL quasars,
the spectral shape cannot be constrained due to the nondetection in the hard band.
We used the Bayesian code {\sc behr} \citep{Park2006} to obtain a best-guess estimate of the effective photon index \citep[see Section~4.4 of][]{Luo2017}, 
which is $\Gamma_{\rm eff} \approx 1.8$.
Given this estimated $\Gamma_{\rm eff}$ and the \hbox{hard-band} flux weakness, 
the stacked source of the five LoBAL quasars is perhaps also intrinsically \xray\ weak.

In order to identify intrinsically \xray\ weak AGNs individually and improve the constraints upon the fraction of such objects among  BAL quasars,
we analyzed additional \chandra\ observations for the seven HiBAL quasars. The HiBAL quasars are the majority population of BAL quasars, and they are also more amenable to economical \chandra\ observations than the LoBAL quasars; thus we focus upon the seven HiBAL quasars in this study.
One object (\hbox{LBQS $0021-0213$}) was serendipitously detected (3.8\arcmin\   away from the observation aim point) in a new \chandra\ archival observation (Cycle~9, observation ID 8918) in addition to the observation in \cite{Gallagher2006}, and it was observed with the \hbox{I-array} of the Advanced CCD Imaging Spectrometer \cite[ACIS,][]{Garmire2003} for a 29.8~ks exposure.  
The other six objects have our new \chandra\ observations in Cycle 16, and they were observed using the S3 CCD of ACIS with $\approx$ 10 ks exposures. 
The details of the \chandra\ observations for the seven objects are listed in Table~\ref{tbl-obs}. Table~\ref{tbl-obs} also includes the BAL properties of these objects adopted from \cite{Gallagher2006}.
The redshift versus apparent and absolute \hbox{{\it B}-band} magnitude distributions for the 35 BAL quasars in the \cite{Gallagher2006} sample (including our subsample of the seven quasars) and objects from the SDSS DR7 quasar catalog \citep{Schneider2010} are shown in Figure \ref{fig-lz}.
Our targets are among the most luminous quasars in the optical/UV. Their high luminosities and redshifts make them more representative of the typically studied \hbox{BAL-quasar} population than the 
generally low-luminosity and \hbox{low-redshift} objects amenable to observations with \nustar\ .

\section{X-RAY DATA ANALYSIS}\label{sec:data}
Each of our seven sample objects has two \chandra\ observations. 
We analyzed the observational data using the \chandra\ Interactive Analysis
of Observations (CIAO; v4.9) tools.
For each observation, 
we ran the {\sc chandra\_repro} script to generate a new level 2 event file, 
and we filtered background flares by 
running the {\sc deflare} script with an iterative 3$\sigma$ clipping algorithm. 
The images in the full (\hbox{0.5--8~keV}), 
 soft (\hbox{0.5--2~keV}), and hard (\hbox{2--8~keV}) bands were constructed by running a {\sc dmcopy} script.
To detect corresponding \xray\ sources for our sample objects and search for other sources,
we used the automated \hbox{source-detection} tool {\sc wavdetect} \citep{Freeman2002} 
with a false-positive probability threshold of $10^{-6}$ and wavelet scale sizes of 1, 1.414, 2, 2.828, 4, 5.656, and 8 pixels. 
Each source is detected in at least one of the two observations,
which contains an \xray\ point source at a location consistent with the optical position of the target.
The measured X-ray-to-optical positional offsets of the objects span a range of 0.2\arcsec--2.0\arcsec\ with a mean value of 0.8\arcsec.

We extracted source counts 
in the three energy bands with aperture photometry. 
The \hbox{source-extraction} region is a circular aperture centered on 
the \xray\ position\footnote{\hbox{LBQS $1212+1445$} and \hbox{LBQS $2201-1834$} are detected only in one of the two observations, and 
the measured \hbox{X-ray-to-optical} positional offsets are 0.6\arcsec and 1.1\arcsec, respectively.
For each of the two sources, we adopted its detected \xray\ position in both observations to extract \xray\ photometry. 
The results do not change if we adopt their optical positions to extract \xray\ photometry.} with a 2\arcsec\ radius.
For the on-axis observations, 
the \hbox{encircled-energy} fractions (EEFs) for the source aperture are 0.939, 0.959, and 0.907 in 
the full, soft, and hard bands, respectively. 
For \hbox{LBQS $0021-0213$} in the second observation with an \hbox{off-axis} angle of 3.8\arcmin, 
the EEFs are 0.812, 0.834, 0.760 in the full, soft, and hard bands, respectively.
The background was estimated over an annulus centered on the same position with a 10\arcsec\ inner radius
and a 40\arcsec\ outer radius.
There are no \xray\ sources in the background region for any observation.
For each band of each source,
we summed the extracted counts from the \hbox{source-extraction} regions of the previous and new observations, to obtain the total source counts ($S$).
Similarly, the total background counts ($B$) were the sum of the extracted counts from the background annulus of the two observations.
The $1\sigma$ errors on the extracted source and background counts were derived following the Poisson approach of \cite{Gehrels1986}.
The net counts were obtained by subtracting from
the source counts 
the estimated number of background counts in the source aperture, which was scaled from the total background counts with a scaling factor ($BACKSCAL$) being the ratio between
the areas of the background and the \hbox{source-extraction} regions.

We then calculated a binomial no-source probability, $P_{\rm B}$, to assess the significance of 
the source signal \citep[e.g.,][]{Broos2007,Xue2011,Luo2013,Luo2015}, which is defined as 
\begin{equation}
P_{\rm B}=\sum_{X=S}^{N}\frac{N!}{X!(N-X)!}p^X(1-p)^{N-X}~,
\end{equation}
where $N=S+B$ and $p=1/(1+BACKSCAL)$. 
$P_{\rm B}$ represents the probability of observing $\ge S$ counts in the source-extraction region under the assumption that there is no real source at the relevant location.
A smaller $P_{\rm B}$ value indicates a more significant signal. 
If the measured $P_{\rm B}$ value in a given band is smaller than a given threshold, 
we considered the source detected in this band, and provided measurements of the photometric properties; otherwise we provided \hbox{upper-limit} constraints.
We adopted a $P_{\rm B}$ threshold of 0.04 (corresponding to a $\approx 2.1\sigma$ significance level in a Gaussian distribution), which is appropriate for extracting \xray\ photometry of sources at \hbox{pre-specified} positions \citep[e.g.,][]{Luo2014,Luo2015}. 
Given this threshold, we expect only 0.28 false detections ($P_{\rm B}<0.04$ caused by background fluctuations) in a given band for seven trials/targets.
Four of the seven targets are significantly ($>4\sigma$) detected in both the soft and hard bands.
Two targets (\hbox{LBQS $1203+1530$} and \hbox{LBQS $1442-0011$}) are significantly ($>4\sigma$) detected in the soft band, while they are weakly detected in the hard band with $P_{\rm B}$ values of 0.033 and 0.019, respectively. 
These relatively weak signals are reflected in the large uncertainties on their \hbox{hard-band} counts.
The other target (\hbox{LBQS $2201-1834$}) is undetected in both the soft and hard bands.
For the detected sources, we provided \hbox{net-count} measurements along with their $1\sigma$ errors; 
the errors were derived from the $1\sigma$ errors on the extracted source
and background counts, using the formula for the propagation of
uncertainties (as in, e.g., Section 1.7.3 of \citealt{Lyons1991}).
For the undetected source, we derived 90\% \hbox{confidence-level} upper limits on the source counts using the Bayesian approach of \cite{Kraft1991}. 
The source counts or their upper limits in the soft and hard bands are listed in Table~\ref{tbl-xbasic}.

The effective photon index ($\Gamma_{\rm eff}$) for a \hbox{power-law} spectrum 
was derived from the observed band ratio, 
which is defined as the ratio between the \hbox{soft-band} and the \hbox{hard-band} counts.
The procedure is as follows:
(1) for each observation, we generated a set of mock spectra using the {\sc fakeit} routine in XSPEC \citep[v12.9.1;][]{Arnaud1996} and the spectral response files,
assuming a set of $\Gamma$ values for a \hbox{power-law} model that is modified by Galactic absorption \cite[e.g.,][]{Gallagher2006,Luo2017};
(2) we then computed the corresponding set of band ratios from the simulated spectra;
(3) the results from the two observations of each source were combined by calculating the \hbox{exposure-time} weighted means of the two sets of band ratios;
(4) the effective photon index was then derived from the observed band ratio by a simple interpolation of the \hbox{$\Gamma$-band} ratio pairs.
We used {\sc behr} to 
derive the $1\sigma$ errors on the band ratios.
The errors on $\Gamma_{\rm eff}$ were propagated from
the errors on the band ratios.

We also used XSPEC to fit the target spectra.
For each observation, the source spectrum was extracted from a circular region centered on the \xray\ position with a radius of 3\arcsec\ for the on-axis objects or a radius of 4\arcsec\ for the off-axis object (\hbox{LBQS $0021-0213$} in the second observation). 
The background spectrum was extracted from an annulus centered on the same position with a 10\arcsec\ inner radius and a 40\arcsec\ outer radius.
For each source, we combined the spectra of two observations. 
We used the C-statistic and a power-law model modified by Galactic absorption ({\sc wabs*zpowerlw}) to fit the 0.5--8~keV combined spectrum. 
Only spectra with more than 4 counts are useful, and thus we did not fit the spectrum of \hbox{LBQS $2201-1834$}.
The total numbers of counts in the combined spectra of the other targets span a range of 7--34.
The best-fit $\Gamma_{\rm XSPEC}$ values and their 1 $\sigma$ errors are listed in column 6 of Table~\ref{tbl-xbasic}.
These $\Gamma_{\rm XSPEC}$ values are in general consistent with the $\Gamma_{\rm eff}$ values, considering their large uncertainties.
In the following analyses and discussions, we adopted the $\Gamma_{\rm eff}$ values; using $\Gamma_{\rm XSPEC}$ instead would not change our results.

We converted the observed source count rates to fluxes using a \hbox{power-law} spectrum model with a photon index of $\Gamma_{\rm eff}$ and the spectral response files.
The errors on the fluxes or flux densities were propagated from the errors on the net counts.
We used the measured $\Gamma_{\rm eff}$ value and the \hbox{soft-band} flux to derive a 2~keV flux density ($f_{\rm2keV,soft}$), and computed the \hbox{soft-band} derived \hbox{optical-to-X-ray} slope ($\alpha_{\rm OX,soft}$).
Additionally, we used $\Gamma=2$ and the \hbox{hard-band} flux to calculate a \hbox{hard-band} derived 2~keV flux density ($f_{\rm2keV,hard}$), and computed the \hbox{hard-band} derived $\alpha_{\rm OX,hard}$ (see Footnote~10).
The flux density at \hbox{rest-frame} $2500~{\textup{\AA}}$ ($f_{\rm 2500~{\textup{\AA}}}$) and the $2500~{\textup{\AA}}$ monochromatic luminosity ($l_{\rm 2500~{\textup{\AA}}}$) used to compute $\alpha_{\rm OX,soft}$ and $\alpha_{\rm OX,hard}$ were adopted from \cite{Gallagher2006}.
The errors of $\alpha_{\rm OX,soft}$ ($\alpha_{\rm OX,hard}$) were propagated from the errors of $f_{\rm 2keV,soft}$ ($f_{\rm 2keV,hard}$).
For the one undetected source (\hbox{LBQS $2201-1834$}), we assumed $\Gamma_{\rm eff}=1$ to calculate the upper limits on fluxes and $f_{\rm 2keV,soft}$ (and subsequently $\alpha_{\rm OX,soft}$) because of its \hbox{BAL-quasar} nature, which suggests a hard spectrum due to absorption \cite[e.g.,][]{Gallagher2006}.
We used $\Gamma=2$ and the \hbox{hard-band} flux upper limit to calculate the upper limit on $f_{\rm 2keV,hard}$ (and subsequently $\alpha_{\rm OX,hard}$).
In Table~\ref{tbl-xbasic}, we list the band ratio, $\Gamma_{\rm eff}$, and flux values for our sample objects. The other \xray\ properties are listed in Table~\ref{tbl-aox}.

\section{TWO GOOD CANDIDATES FOR INTRINSICALLY X-RAY WEAK AGNS}\label{sec:result}

\begin{figure}
\centerline{
\includegraphics[scale=0.42]{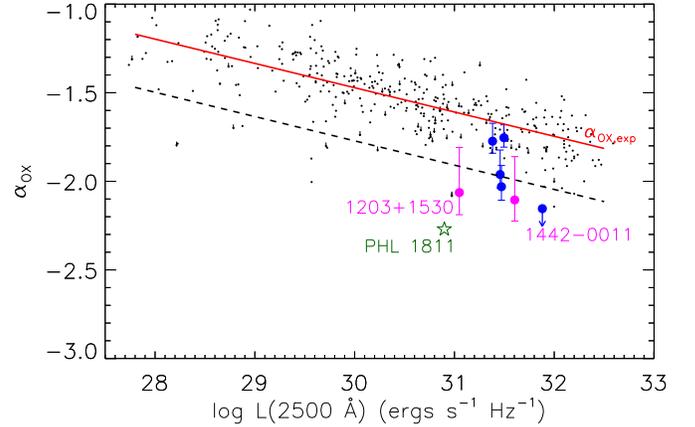}
}
\caption{
X-ray-to-optical power-law slope ($\alpha_{\rm OX}$) 
vs.~$2500~{\textup{\AA}}$ 
monochromatic luminosity for our seven sample objects (\hbox{LBQS $1203+1530$} and \hbox{LBQS $1442-0011$} in magenta and the others in blue).
The small black dots and downward arrows (upper limits) are 
for the typical AGN samples in \citet{Steffen2006},
and the solid red line shows the \hbox{$\alpha_{\rm OX}$--$L_{\rm 2500~{\textup{\AA}}}$} relation. 
PHL 1811 is shown as a green star.
For our sample objects, the \hbox{$y$-axis} values ($\alpha_{\rm OX}$) 
were computed using the \hbox{hard-band} fluxes and $\Gamma = 2$.
Objects lying below the dashed line ($\Delta\alpha_{\rm OX}=-0.3$) show significant \hbox{hard-band} flux weakness.
}
\label{fig-aox-l2500}
\end{figure}

\begin{figure}
\centerline{
\includegraphics[scale=0.42]{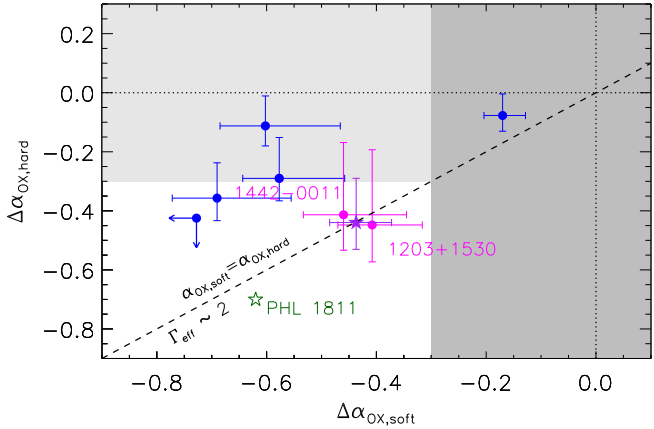}
}
\caption{An updated version of Figure~\ref{fig-aox-old},
showing $\Delta\alpha_{\rm OX,hard}$ versus $\Delta\alpha_{\rm OX,soft}$
for our seven sample objects (blue and magenta data points) and
PHL 1811 (green star) after considering the new observations.
\hbox{LBQS $1203+1530$} and \hbox{LBQS $1442-0011$} are shown as magenta data points, and
the stacked source of these two objects is shown as a purple star.
For the detected objects, the uncertainties of $\Delta\alpha_{\rm OX,hard}$ ($\Delta\alpha_{\rm OX,soft}$) are shown.
For the single object (\hbox{LBQS $2201-1834$}) that is undetected in both the soft and hard bands, upper limits on $\Delta\alpha_{\rm OX,hard}$ ($\Delta\alpha_{\rm OX,soft}$) are shown as arrows.
The dark and light shaded regions show the $\approx 95\%$ ($\approx 2\sigma$) confidence-level uncertainties of $\alpha_{\rm OX,exp}$.
The slanted dashed line indicates
$\alpha_{\rm OX,soft}$ = $\alpha_{\rm OX,hard}$ and $\Gamma_{\rm eff}\approx 2$.
Objects lying outside the shaded regions are significantly \xray\ weak,
and objects lying near the slanted line have soft spectra.
}
\label{fig-daox}
\end{figure}

The seven sample objects were undetected 
in the hard band (2--8 keV, $\approx$ 6--24 keV in the rest frame)
in the previous $5\textrm{--}7~\rm ks$ \chandra\ observations of \cite{Gallagher2006}.
After combining the data from previous and new observations,
the exposure times reach 13.6--36.6~ks.
Except for one object (\hbox{LBQS $2201-1834$}) that is still undetected in the hard band,
four objects are now significantly ($> 3\sigma$) detected,
and the other two objects (\hbox{LBQS $1203+1530$} and \hbox{LBQS $1442-0011$}) are weakly detected ($2.1 \sigma$ and $2.3 \sigma$, respectively).
Half of the six detected objects show hard \xray\ spectral shapes with $\Gamma_{\rm eff}$ values in the range of $0.4\textrm{--}1.1$ ($-0.4\textrm{--}1.5$ considering the 1 $\sigma$ uncertainties).
The others (including \hbox{LBQS $1203+1530$} and \hbox{LBQS $1442-0011$}) 
show soft \xray\ spectral shapes with $\Gamma_{\rm eff} > 1.7$.

Figure 3 shows the $\alpha_{\rm OX}$ versus $L_{\rm 2500~{\textup{\AA}}}$ distribution for our seven sample objects, PHL 1811, and the
typical AGN samples in \citet{Steffen2006}.
For our sample objects, we used the \hbox{hard-band} derived $\alpha_{\rm OX,hard}$ values.
We considered $\Delta\alpha_{\rm OX,hard} < -0.3$ (corresponding to a \hbox{hard-band} flux weakness factor \footnote{It is calculated as $f_{\rm weak}= 10^{ -\Delta\alpha_{\rm OX,hard}/0.3838}$.} of $f_{\rm weak}>6$) as the criterion for a object being X-ray weak (see Section~\ref{sec:sample}), indicated by the dashed line in Figure~\ref{fig-aox-l2500}. 
Four objects in our sample (including the undetected object) meet this criterion, showing significant \hbox{hard-band} flux weakness.

We applied the method introduced in Section~\ref{sec:sample} to identify intrinsically \xray\ weak AGNs in our sample.
Similar to Figure~\ref{fig-aox-old},
Figure~\ref{fig-daox} shows a comparison of the \hbox{soft-band} flux weakness ($\Delta\alpha_{\rm OX,soft}$) to the \hbox{hard-band} flux weakness ($\Delta\alpha_{\rm OX,hard}$)
for the seven sample objects.
Compared to Figure~\ref{fig-aox-old}, the \hbox{hard-band} flux weakness of six objects can be constrained now.
In particular, two sources (\hbox{LBQS $1203+1530$} and \hbox{LBQS $1442-0011$}, magenta points in Figures~\ref{fig-aox-l2500} and \ref{fig-daox})
 lie outside the shaded regions and close to the slanted line, showing significant \hbox{hard-band} flux weakness and soft spectral shapes.
For \hbox{LBQS $1203+1530$} (\hbox{LBQS $1442-0011$}),
the effective power-law photon index is
$\Gamma_{\rm eff} = 2.2^{+0.9}_{-0.9}$ 
($\Gamma_{\rm eff} =1.9_{-0.8}^{+0.9}$), 
and the factor of \hbox{hard-band} flux weakness ($f_{\rm weak}$) is 14.7 (11.9). 
Therefore, we suggest that these two sources are good candidates for intrinsically \xray\ weak AGNs.
We note that due to limited photon statistics, there are substantial uncertainties on the derived effective photon indices and factors of \hbox{hard-band} flux weakness for \hbox{LBQS $1203+1530$} and \hbox{LBQS $1442-0011$}; e.g., their $\alpha_{\rm OX,hard}$ error bars extend beyond the $\Delta\alpha_{\rm OX}=-0.3$ line in Figure~\ref{fig-aox-l2500}, and the upper boundaries correspond to 
$f_{\rm weak}$ of only 3.2 and 2.8, respectively.  
Deeper observations are needed to constrain better their
spectral shapes and \hbox{hard-band} flux weakness, and confirm their intrinsically \xray\ weak nature.

We also stacked the \xray\ emission of \hbox{LBQS $1203+1530$} and \hbox{LBQS $1442-0011$}, aiming to obtain a more significant detection and constrain their average \xray\ properties.
Indeed, the stacked source is significantly ($3.1\sigma$) detected in the hard band , and it also has significant \hbox{hard-band} flux weakness ($f_{\rm weak}=14.0$) and a soft spectral shape ($\Gamma_{\rm eff} = 2.1^{+0.6}_{-0.6}$).
The stacked source is shown as a purple star in Figure~\ref{fig-daox},
which has smaller errors of $\Delta\alpha_{\rm OX,soft}$ and $\Delta\alpha_{\rm OX,hard}$
compared to the two objects individually.

One of the seven targets, \hbox{LBQS $2201-2834$}, 
remains undetected in the soft and hard bands.
It is significantly \xray\ weak in the hard band ($f_{\rm weak}>$12.8),
but we cannot constrain its spectral shape due to the nondetection.
It could still be an intrinsically \xray\ weak AGN, 
and a deeper \xray\ observation is required to determine its nature.

\subsection{Fraction of Intrinsically X-ray Weak BAL Quasars}\label{sec:frac}
The \cite{Gallagher2006} LBQS BAL quasars are the only \hbox{well-defined} \hbox{BAL-quasar} sample 
that has been investigated systematically for the presence of intrinsically \xray\ weak AGNs.
With the two good candidates identified in this study,
we can constrain the fraction of intrinsically \xray\ weak AGNs. 
In addition to our seven sample objects, the other 22 objects among the 29 HiBAL quasars in \cite{Gallagher2006} do not show evidence of intrinsic \xray\ weakness \cite[see Section 4.2.3 of ][]{Luo2013}.
Since the undetected source in our sample (\hbox{LBQS $2201-1834$}) could still be a candidate intrinsically \xray\ weak AGN, 
we consider there to be 2--3 candidates for intrinsically \xray\ weak AGNs. 
We have thus constrained the fraction of intrinsically \xray\ weak AGNs among HiBAL quasars to be $\approx 7^{+8}_{-2}\textrm{--}10^{+9}_{-3}\%$ (\hbox{2/29--3/29});
the $1 \sigma$ Poisson uncertainties were computed following the approach in \cite{Cameron2011}.  

In addition to the 29 HiBAL quasars, there are six LoBAL quasars in the sample of \cite{Gallagher2006}. 
Five of the six LoBAL quasars are undetected in the hard band, 
and the stacked upper limit in Figure~\ref{fig-aox-old} suggests the presence of intrinsically \xray\ weak AGNs among these five objects (see also Section~\ref{sec:sample}). 
There is probably a substantial fraction of intrinsically \xray\ weak AGNs among the five \hbox{hard-band} undetected LoBAL quasars based on the following considerations: 
(1) Among the 29 HiBAL quasars, 
the three candidates for intrinsically \xray\ weak AGNs 
(\hbox{LBQS $1203+1530$}, \hbox{LBQS $1442-0011$}, and \hbox{LBQS $2201-1834$}) 
are the \xray\ weakest in the hard band (see Figure~9 of \citealt{Luo2013} and Figure~\ref{fig-daox});
(2) Probably intrinsically \xray\ weak BAL quasars 
tend to be \xray\ weaker than \xray\ absorbed BAL quasars;
(3) Given the individual and stacked upper limits on $\Delta\alpha_{\rm OX,hard}$ of the five hard-band undetected LoBAL quasars, they appear to be \xray\ weaker than the HiBAL quasars in general (Figure~\ref{fig-aox-old}).
Therefore, it is probable that the fraction of intrinsically \xray\ weak AGNs among LoBAL quasars is larger than that in HiBAL quasars.
If all the five \hbox{hard-band} undetected LoBAL quasars are intrinsically \xray\ weak,
there would be up to 8 intrinsically \xray\ weak AGNs 
among the 35 BAL quasars in \cite{Gallagher2006}. 
Therefore, the fraction of intrinsically \xray\ weak AGNs among the entire BAL quasar population (both HiBAL and LoBAL quasars) is $\approx 6^{+6}_{-2}\textrm{--} 23^{+8}_{-6}\%$ (2/35--8/35).
This is a more robust constraint compared to the $\approx 17\textrm{--}40\%$ fraction suggested in \cite{Luo2013}, 
which was estimated solely from the stacking analyses.

\section{DISCUSSION}\label{sec:discuss}
\begin{figure*}
\centerline{
\includegraphics[width=\textwidth]{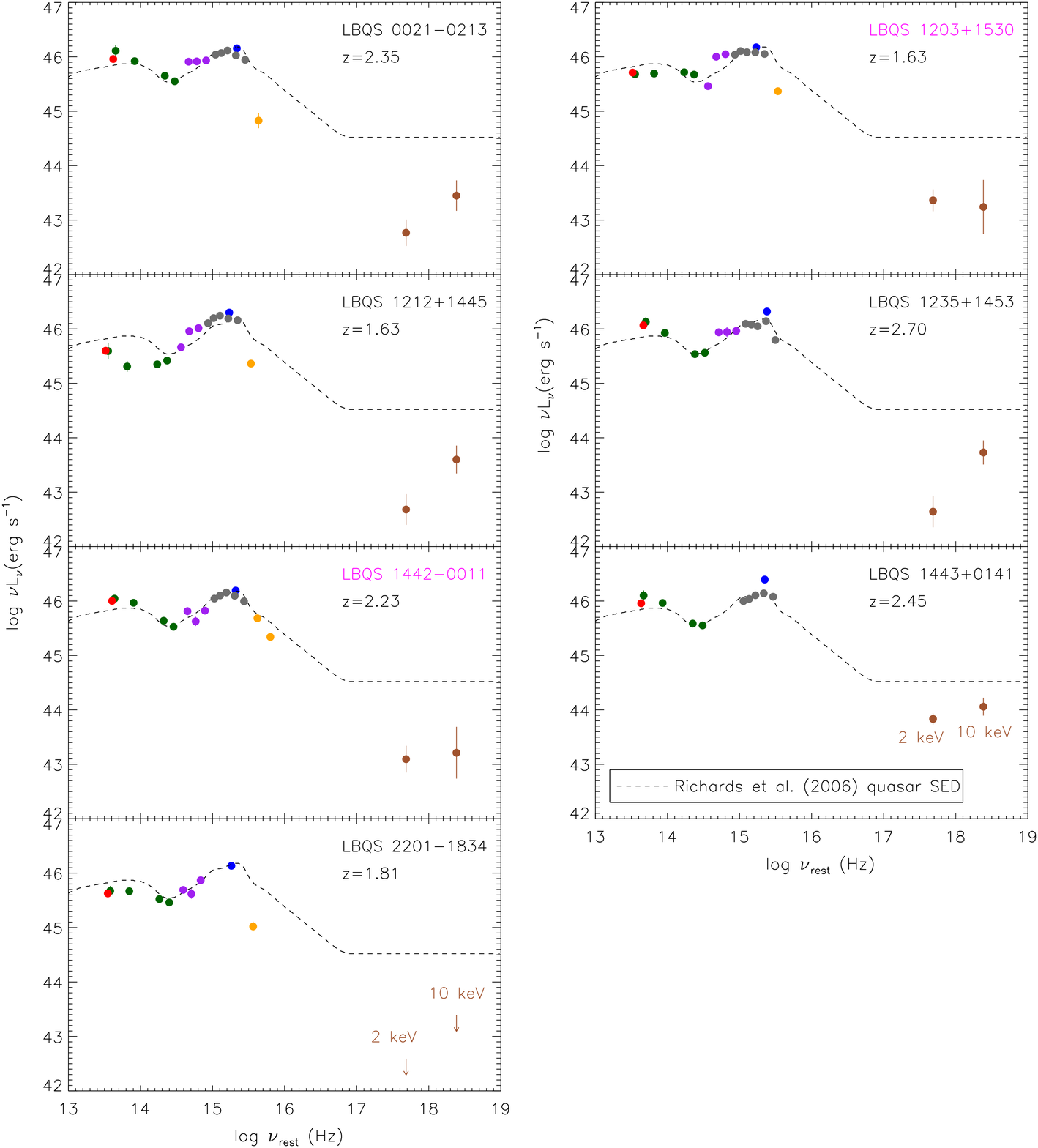}
}
\caption{Rest-frame IR-to-X-ray SEDs for the seven sample objects. The IR-to-UV photometric data points were gathered from {\it Spitzer} MIPS (24~$\rm \mu m$, red), {\it WISE} (green), 2MASS  (purple), SDSS (gray), LBQS (blue), and {\it GALEX} (orange) catalogs. The 2~keV and 10~keV data of \chandra\ (brown points and arrows) were derived from 0.5--2~keV and 2--8~keV fluxes (or flux upper limits), respectively. The SED for each object was scaled to the composite quasar SED of optically luminous SDSS quasars \citep{Richards2006} at \hbox{rest-frame} $1~\rm \mu m$.
The object names (\hbox{LBQS $1203-1530$} and \hbox{LBQS $1442-0011$}) of our two good candidates for intrinsically \xray\ weak AGNs are highlighted in magenta. 
These \hbox{multi-band} observations are non-simultaneous, and thus the SEDs may be affected by variability.
}
\label{fig-sed}
\end{figure*}

\begin{figure*}
 \centerline{
  \includegraphics[scale=0.42]{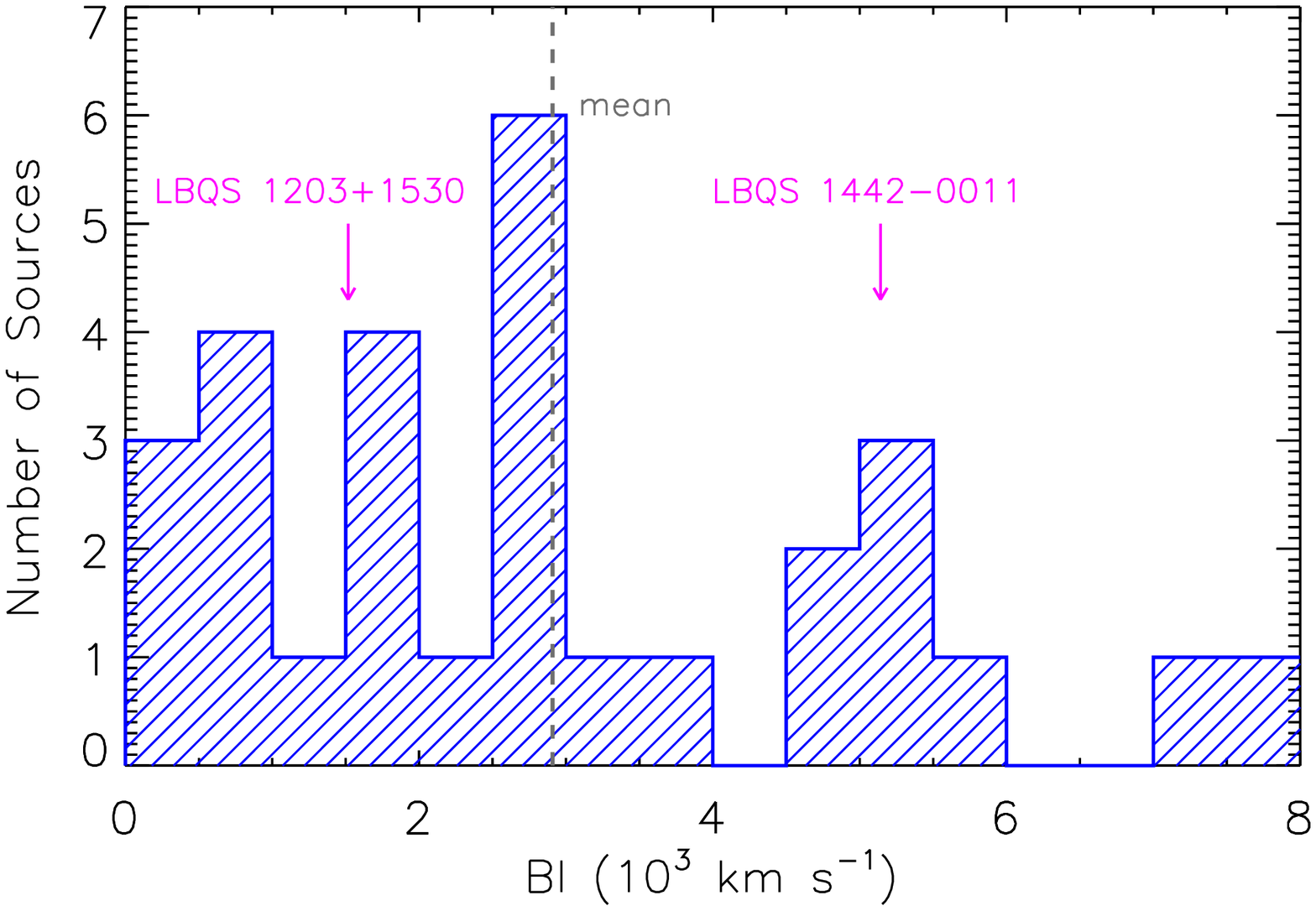}
  \includegraphics[scale=0.42]{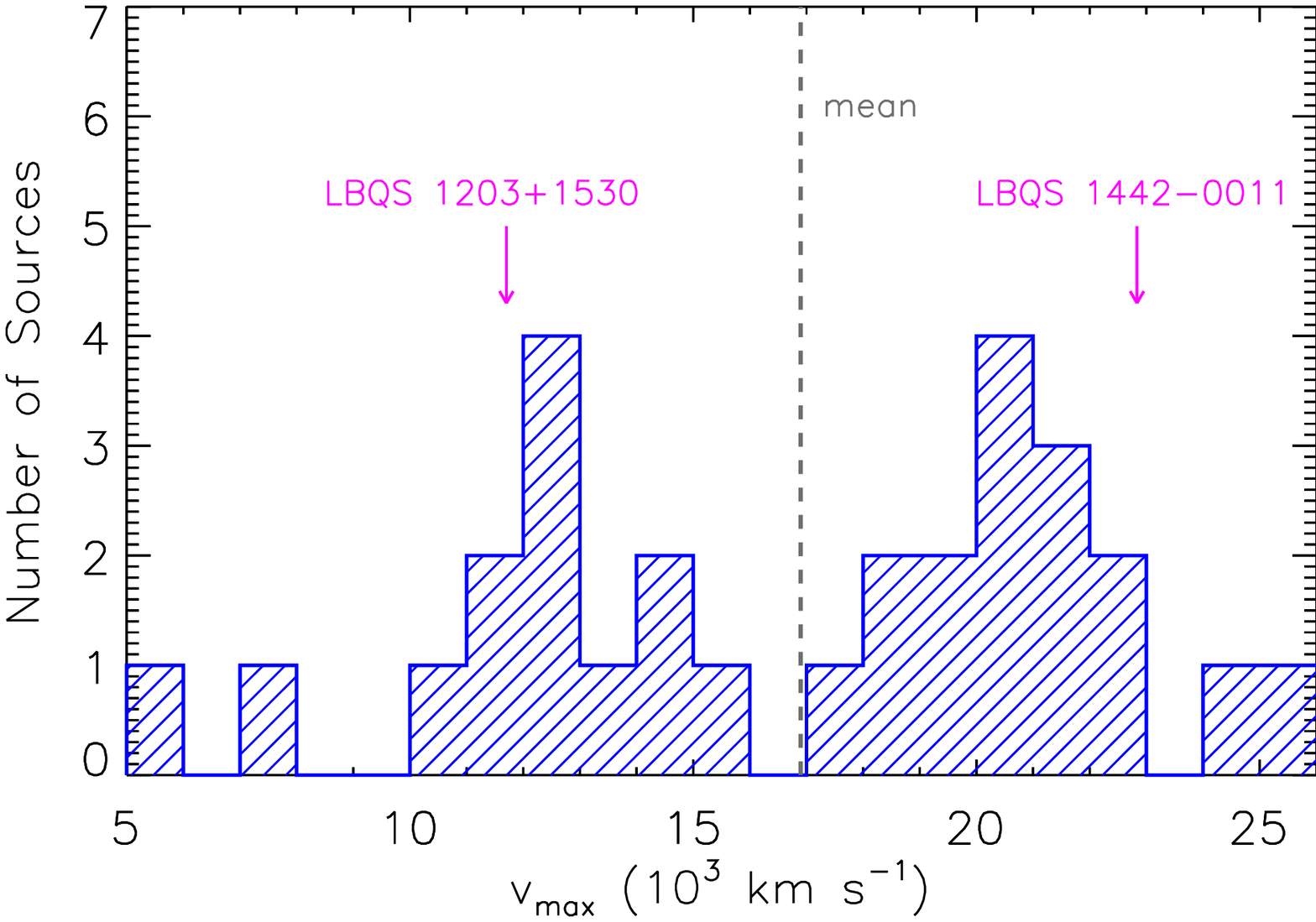}
}
\caption{Distribution of (a) BALnicity index (BI) and (b) maximum velocity of \ion{C}{4} absorption blueshift ($v_{\rm max}$) for the 29 HiBAL quasars in the \cite{Gallagher2006} sample (blue histogram). The vertical dashed line marks the mean value of each  parameter for the 29 objects. The magenta arrows indicate BI and $v_{\rm max}$ values for \hbox{LBQS $1203+1530$} and \hbox{LBQS $1442-0011$}. 
 \label{fig-BAL}} 
\end{figure*}

As discussed in Section~\ref{sec:sample}, a soft spectrum in the hard \xray\ ($\ga$~8~keV) band indicated by a steep effective \hbox{power-law} photon index can rule out the general \hbox{Compton-thick} absorption scenario. 
However, there exists an alternative \hbox{Compton-thick} scenario with an exceptional geometrical configuration \cite[e.g.,][]{Murphy2009,Matt2012,Bianchi2017}. 
In this scenario, 
there is a very compact neutral Compton-thick absorber 
  with an extremely high column density ($N_{\rm H} \ga 10^{25} \rm cm^{-2}$) along the line of sight. 
The intrinsic emission along the \hbox{line-of-sight} is completely absorbed up to very high energies, 
and there is hardly any \hbox{Compton-reflection} hump emerging in the spectrum due to the very small covering factor of the neutral absorber to the \xray\ corona.
Thus the observed spectrum is dominated by a soft reflected component
from a \hbox{large-scale} highly ionized "mirror" up to hard \hbox{X-rays} ($\approx 8\textrm{--}20~\rm keV$).
The scenario requires a rather peculiar and finely tuned configuration, 
and there have been no such objects discovered before, and thus we acknowledge such a possibility, but consider it unlikely.

We constructed \hbox{rest-frame} infrared (IR) to \xray\ spectral energy distributions (SEDs) for our seven sample objects, as shown in Figure~\ref{fig-sed}. 
The \hbox{IR-to-UV} photometric data were gathered from the public catalogs of Multiband Imaging Photometer for {\it Spitzer} \cite[MIPS;][]{Rieke2004}, {\it Wide-field~Infrared~Survey~Explorer} \cite[{\it WISE};][]{Wright2010}, Two Micron All Sky Survey \cite[2MASS;][]{Skrutskie2006}, Sloan Digital Sky Survey \cite[SDSS;][]{York2000},
and {\it Galaxy~Evolution~Explorer} \cite[{\it GALEX};][]{Martin2005}. 
The optical and UV data have been corrected for Galactic extinction following the dereddening approach of \cite{Cardelli1989} and \cite{Donnell1994}. 
We also used the \hbox{$B_{\rm J}$-band} magnitude adopted from \citet{Gallagher2006} and the 2~keV and 10~keV luminosities, 
which were derived from the soft and hard bands, respectively,
by assuming a \hbox{power-law} model with measured $\Gamma_{\rm eff}$.
These SEDs are in general consistent with those presented in \cite{Gallagher2007}. 
For comparison, the composite SED of typical SDSS quasars in \cite{Richards2006} is also shown in Figure~\ref{fig-sed}.
Most of the \hbox{mid-infrared-to-UV} SEDs of these objects are generally consistent with those of typical quasars, 
except for one peculiar object (\hbox{LBQS $1212+1445$}), 
which shows unusually weak \hbox{mid-infrared-to-near-infrared} emission (see the {\it WISE} data points in the corresponding panel of Figure~\ref{fig-sed}), 
suggesting that it could be a "hot-dust-poor quasar" (e.g., \citealt{Jiang2010,Hao2011,Lyu2017}). 
Compared to typical quasars,
 \hbox{LBQS $1203+1530$} and \hbox{LBQS $1442-0011$}
 have weak \xray\ emission, 
 and the two \xray\ data points of these two objects also indicate a steep spectral shape ($\Gamma \approx 2$) that is typical of type~1 quasars.
 
We have estimated that the fraction of intrinsically \xray\ weak AGNs among BAL quasars is $\approx 6^{+6}_{-2}\textrm{--} 23^{+8}_{-6}\%$.
This fraction appears considerably larger than the $\lesssim~2\%$ fraction among \hbox{non-BAL} quasars \citep[e.g.,][]{Gibson2008a}.
One possible interpretation of this significant difference is that 
disk winds in intrinsically \xray\ weak BAL quasars can be launched easily with larger covering factors of the nuclei,
as the \hbox{line-driven} winds are not significantly ionized by the nuclear \xray\ emission;
thus intrinsically X-ray weak quasars would be preferentially observed as BAL quasars \cite[e.g.,][]{Luo2013}.
We investigate whether \hbox{LBQS $1203+1530$} and \hbox{LBQS $1442-0011$} show obviously stronger outflowing winds,
by comparing their BALnicity Indices \citep[BI; the modified  equivalent width of the BAL as defined in][]{Weymann1991} and maximum velocities of \ion{C}{4} absorption blueshift ($v_{\rm max}$) to those of the other HiBAL quasars in \cite{Gallagher2006}.
The distributions of the two parameters are shown in Figure~\ref{fig-BAL}.
\hbox{LBQS $1442-0011$} shows relatively large BI and $v_{\rm max}$ values among these HiBAL quasars, 
but \hbox{LBQS $1203+1530$} does not show these characteristics. 
It is probable that the two parameters alone are not good indicators of wind strength due to the complication from the orientation effect.
Another interpretation is that disk winds in intrinsically \xray\ weak BAL quasars are not necessarily extremely strong (with high column densities and/or large velocities) 
but just have large covering factors.

We could also estimate the frequency of intrinsically \xray\ weak AGNs among the general type~1 quasar population 
and assess its effect on \xray\ surveys for AGN census work.
The fraction of intrinsically \xray\ weak AGNs among HiBAL quasars that we constrained is $\approx 7\textrm{--}10\%$,
and the fraction of intrinsically \xray\ weak AGNs among BAL quasars could be up to $\approx 23\%$ 
(see Section~4.1).
Since the subclass of BAL quasars is thought to account for $\approx 15\%$ of quasars \citep[e.g,][]{Hewett2003,Reichard2003,Gibson2009,Trump2006,Allen2011},
the fraction of intrinsically \xray\ weak BAL quasars among the general quasar population should be $\lesssim 3.5\%$.
In addition, the fraction of intrinsically \xray\ weak AGNs among non-BAL quasars is $\lesssim 2\%$ \citep{Gibson2008a},
which translates to a $\lesssim 1.7\%$ fraction among the general 
type~1 quasar population.
In total, intrinsically \xray\ weak AGNs likely comprise a small minority ($\lesssim 5.2\%$) of the luminous type~1 AGN population, 
and they should not affect significantly the completeness of these AGNs found in deep \xray\ surveys.
This is consistent with AGN selection results for sensitive multiwavelength survey fields \citep[e.g.,][]{Brandt2015}.

The underlying physics responsible for the intrinsically weak \xray\ emission remains unclear. It could be related to some process that quenches the coronal \xray\ emission \citep{Proga2005,Leighly2007b,Luo2013}.
Another possible scenario is that intrinsic \xray\ weakness is related to a very high accretion rate of the accretion flow \citep[e.g.,][]{Leighly2007b,Luo2014}, 
where the accretion timescale is shorter than the diffusion timescale of \xray\ photons that are produced close to the BH. 
We thus searched for \hbox{Eddington-ratio} estimates ($L/L_{\rm Edd}$) for the 29 HiBAL quasars in \cite{Gallagher2006} from the literature, 
 and 19 of them have such estimates from \cite{Yuan2003}, \cite{Dietrich2009} and \cite{Shen2011}.   
\hbox{LBQS $1442-0011$} and \hbox{LBQS $1203+1530$} do not show high Eddington ratios (0.25 and 0.17, respectively) relative to the other HiBAL quasars with Eddington ratios in the range of 0.14\textrm{--}1.47.
However, there are only six sources (including \hbox{LBQS $1442-0011$}) with BH-mass estimates based on the $\rm H \beta$ line profile using the single-epoch virial mass approach, and the other objects (including \hbox{LBQS $1203+1530$}) have only \ion{Mg}{2}- or \hbox{\ion{C}{4}-based} \hbox{BH-mass} estimates \citep{Shen2011}, that are less reliable or even systematically in error.
Additional \hbox{near-IR} spectra are required to provide more reliable $\rm H \beta$-based estimates of BH masses and Eddington ratios, and assess if intrinsic \xray\ weakness is related to very high accretion rates.

\section{SUMMARY AND FUTURE WORK}\label{sec:sum}
In this paper, we have analyzed the combined $\approx 14\textrm{--}37$~ks \chandra\ data of seven \hbox{hard-band} undetected optically bright HiBAL quasars from \cite{Gallagher2006}.
Except for one target that is still undetected,
the other six targets are now detected in the hard band.
We constrained their \hbox{hard-band} (\hbox{rest-frame} $\approx 6\textrm{--}24$ keV) flux  weakness and effective \hbox{power-law} photon indices, 
and found that two targets (\hbox{LBQS $1203+1530$} and \hbox{LBQS $1442-0011$})
show soft spectral shapes ($\Gamma_{\rm eff}= 2.2^{+0.9}_{-0.9}$ and $1.9_{-0.8}^{+0.9}$) and significant \hbox{hard-band} flux weakness (by factors of 14.7 and 11.9), suggestive of being good candidates for intrinsically \xray\ weak BAL quasars. 
These two objects are among the best candidates for intrinsically \xray\ weak AGNs besides PHL 1811.

The 35 LBQS BAL quasars in \cite{Gallagher2006} are the only \hbox{well-defined} \hbox{BAL-quasar} sample 
that has been investigated systematically for the presence of intrinsically \xray\ weak AGNs.
Combined with the results of \cite{Luo2013}, 
we constrained the fraction of intrinsically \xray\ weak AGNs among HiBAL quasars to be $\approx 7 \textrm{--}10\%$ (2/29--3/29).
Since the five \hbox{hard-band} undetected LoBAL quasars in \cite{Gallagher2006} could still be intrinsically \xray\ weak, 
the fraction of intrinsically \xray\ weak AGNs 
among the full BAL quasar population could be $\approx 6\textrm{--} 23\%$ (2/35--8/35). 
This fraction is considerably larger than the fraction ($\lesssim 2\%$) of intrinsically \xray\ weak AGNs among \hbox{non-BAL} quasars, 
suggesting that intrinsically \xray\ weak AGNs are preferentially observed as BAL quasars, probably related to winds launched by accretion disks with larger covering factors of the nuclei. 

Based on the current \chandra\ observational data, 
\hbox{LBQS $1203+1530$} and \hbox{LBQS $1442-0011$} are only weakly detected in the hard band.
In both cases, there are only two counts in the source apertures, 
and the binomial \hbox{no-source} probabilities are $P_{\rm B} \approx$ 0.033 and 0.019 
(corresponding to $2.1\sigma$ and $2.3\sigma$ detections) for \hbox{LBQS $1203+1530$} and \hbox{LBQS $1442-0011$}, respectively.
Thus, there are considerable uncertainties on 
the effective \hbox{power-law} photon indices ($\Gamma_{\rm eff}$) 
and the factors of \hbox{hard-band} \xray\ weakness ($f_{\rm weak}$),  
as shown in Figure~\ref{fig-daox}, Table~\ref{tbl-xbasic}, and Table~\ref{tbl-aox}.
Additional \chandra\ or \hbox{$XMM$-$Newton$} observations are required
in order to improve the detection significance, better constrain the spectral shapes and \hbox{hard-band} flux weakness, 
and confirm the nature of these two objects as intrinsically \xray\ weak AGNs. 
Moreover, \hbox{LBQS $2201-1834$} is still undetected in the soft and hard bands. 
A deeper \xray\ observation is required to constrain its \xray\ spectral shape and determine if it is an intrinsically \xray\ weak AGN. 
This will also help to better constrain the fraction of intrinsically \xray\ weak BAL quasars.

It will also be valuable if deeper observations are performed for the five \hbox{hard-band} undetected LoBAL quasars in \cite{Gallagher2006} to obtain \hbox{hard-band} detections 
and constrain their nature.
These five LoBAL quasars probably host a larger fraction of intrinsically \xray\ weak AGNs. 
However, given their significant \xray\ weakness, 
such observations would be more expensive than those for the HiBAL quasars here.
For example, given the photometric properties for 
the stacked source of the five LoBAL quasars (see Section~\ref{sec:sample} and Figure~\ref{fig-aox-old}),
we estimate that the total \chandra\ exposure time required to obtain a hard-band detection is $\approx 44~\rm ks$ on average.
Therefore,
in order to detect all five LoBAL quasars individually,
we would need to obtain a $\ga 40~\rm ks$ \chandra\ observation for each object in addition to its current $\approx 5~\rm ks$ \chandra\ exposure.

~\\

We thank the referee for reviewing the manuscript carefully and 
providing helpful comments.
We thank Yong~Shi, Qiusheng~Gu, Peng~Wei, Chen~Hu, and Pu~Du
for helpful discussions. 
We acknowledge financial support from
the National Natural Science Foundation of China
grant 11673010 (H.L., B.L.),
National Key R\&D Program of China grant 2016YFA0400702 (H.L., B.L.),
National Thousand Young Talents program of China (B.L.).
W.N.B acknowledges support from 
Chandra X-ray Center grant GO5-16089X, the
NASA ADP Program, and the Penn State ACIS Instrument Team Contract
SV4-74018 (issued by the Chandra X-ray Center, which is operated by the
Smithsonian Astrophysical Observatory for and on behalf of NASA under
contract NAS8-03060). 
S.C.G thanks the Discovery Grant Program of the Natural Science and Engineering Research Council of Canada.

The Guaranteed Time Observations (GTO) for the quasars
studied were selected by the ACIS Instrument Principal
Investigator, Gordon P. Garmire, currently of the Huntingdon Institute
for X-ray Astronomy, LLC, which is under contract to the Smithsonian
Astrophysical Observatory; Contract SV2-82024.

\bibliographystyle{aasjournal}
\bibliography{ms}

\begin{thebibliography}{}
\expandafter\ifx\csname natexlab\endcsname\relax\def\natexlab#1{#1}\fi
\providecommand{\url}[1]{\href{#1}{#1}}

\bibitem[{{Allen} {et~al.}(2011){Allen}, {Hewett}, {Maddox}, {Richards}, \&
  {Belokurov}}]{Allen2011}
{Allen}, J.~T., {Hewett}, P.~C., {Maddox}, N., {Richards}, G.~T., \&
  {Belokurov}, V. 2011, \mnras, 410, 860

\bibitem[{{Arnaud}(1996)}]{Arnaud1996}
{Arnaud}, K.~A. 1996, in Astronomical Society of the Pacific Conference Series,
  Vol. 101, Astronomical Data Analysis Software and Systems V, ed. G.~H.
  {Jacoby} \& J.~{Barnes}, 17

\bibitem[{{Baskin} {et~al.}(2014){Baskin}, {Laor}, \& {Stern}}]{Baskin2014}
{Baskin}, A., {Laor}, A., \& {Stern}, J. 2014, \mnras, 445, 3025

\bibitem[{{Bianchi} {et~al.}(2017){Bianchi}, {Marinucci}, {Matt}, {Middei},
  {Barcons}, {Bassani}, {Carrera}, {La Franca}, \& {Panessa}}]{Bianchi2017}
{Bianchi}, S., {Marinucci}, A., {Matt}, G., {et~al.} 2017, \mnras, 468, 2740

\bibitem[{{Brandt} \& {Alexander}(2015)}]{Brandt2015}
{Brandt}, W.~N., \& {Alexander}, D.~M. 2015, \aapr, 23, 1

\bibitem[{{Broos} {et~al.}(2007){Broos}, {Feigelson}, {Townsley}, {Getman},
  {Wang}, {Garmire}, {Jiang}, \& {Tsuboi}}]{Broos2007}
{Broos}, P.~S., {Feigelson}, E.~D., {Townsley}, L.~K., {et~al.} 2007, \apjs,
  169, 353

\bibitem[{{Cameron}(2011)}]{Cameron2011}
{Cameron}, E. 2011, PASA, 28, 128

\bibitem[{{Cardelli} {et~al.}(1989){Cardelli}, {Clayton}, \&
  {Mathis}}]{Cardelli1989}
{Cardelli}, J.~A., {Clayton}, G.~C., \& {Mathis}, J.~S. 1989, \apj, 345, 245

\bibitem[{{Comastri}(2004)}]{Comastri2004}
{Comastri}, A. 2004, in Astrophysics and Space Science Library, Vol. 308,
  Supermassive Black Holes in the Distant Universe, ed. A.~J. {Barger}, 245

\bibitem[{{Comastri} {et~al.}(2011){Comastri}, {Ranalli}, {Iwasawa}, {Vignali},
  {Gilli}, {Georgantopoulos}, {Barcons}, {Brandt}, {Brunner}, {Brusa},
  {Cappelluti}, {Carrera}, {Civano}, {Fiore}, {Hasinger}, {Mainieri},
  {Merloni}, {Nicastro}, {Paolillo}, {Puccetti}, {Rosati}, {Silverman},
  {Tozzi}, {Zamorani}, {Balestra}, {Bauer}, {Luo}, \& {Xue}}]{Comastri2011}
{Comastri}, A., {Ranalli}, P., {Iwasawa}, K., {et~al.} 2011, \aap, 526, L9

\bibitem[{{Dickey} \& {Lockman}(1990)}]{Dickey1990}
{Dickey}, J.~M., \& {Lockman}, F.~J. 1990, \araa, 28, 215

\bibitem[{{Dietrich} {et~al.}(2009){Dietrich}, {Mathur}, {Grupe}, \&
  {Komossa}}]{Dietrich2009}
{Dietrich}, M., {Mathur}, S., {Grupe}, D., \& {Komossa}, S. 2009, \apj, 696,
  1998

\bibitem[{{Done}(2010)}]{Done2010}
{Done}, C. 2010, ArXiv e-prints, arXiv:1008.2287

\bibitem[{{Fabian} {et~al.}(2017){Fabian}, {Alston}, {Cackett}, {Kara},
  {Uttley}, \& {Wilkins}}]{Fabian2017}
{Fabian}, A.~C., {Alston}, W.~N., {Cackett}, E.~M., {et~al.} 2017,
  Astronomische Nachrichten, 338, 269

\bibitem[{{Fan} {et~al.}(2009){Fan}, {Wang}, {Wang}, {Wang}, {Dong}, {Zhang},
  \& {Cheng}}]{Fan2009}
{Fan}, L.~L., {Wang}, H.~Y., {Wang}, T., {et~al.} 2009, \apj, 690, 1006

\bibitem[{{Freeman} {et~al.}(2002){Freeman}, {Kashyap}, {Rosner}, \&
  {Lamb}}]{Freeman2002}
{Freeman}, P.~E., {Kashyap}, V., {Rosner}, R., \& {Lamb}, D.~Q. 2002, \apjs,
  138, 185

\bibitem[{{Gallagher} {et~al.}(2002){Gallagher}, {Brandt}, {Chartas}, \&
  {Garmire}}]{Gallagher2002a}
{Gallagher}, S.~C., {Brandt}, W.~N., {Chartas}, G., \& {Garmire}, G.~P. 2002,
  \apj, 567, 37

\bibitem[{{Gallagher} {et~al.}(2006){Gallagher}, {Brandt}, {Chartas},
  {Priddey}, {Garmire}, \& {Sambruna}}]{Gallagher2006}
{Gallagher}, S.~C., {Brandt}, W.~N., {Chartas}, G., {et~al.} 2006, \apj, 644,
  709

\bibitem[{{Gallagher} {et~al.}(2007){Gallagher}, {Hines}, {Blaylock},
  {Priddey}, {Brandt}, \& {Egami}}]{Gallagher2007}
{Gallagher}, S.~C., {Hines}, D.~C., {Blaylock}, M., {et~al.} 2007, \apj, 665,
  157

\bibitem[{{Gandhi} {et~al.}(2014){Gandhi}, {Lansbury}, {Alexander}, {Stern},
  {Ar{\'e}valo}, {Ballantyne}, {Balokovi{\'c}}, {Bauer}, {Boggs}, {Brandt},
  {Brightman}, {Christensen}, {Comastri}, {Craig}, {Del Moro}, {Elvis},
  {Fabian}, {Hailey}, {Harrison}, {Hickox}, {Koss}, {LaMassa}, {Luo},
  {Madejski}, {Ptak}, {Puccetti}, {Teng}, {Urry}, {Walton}, \&
  {Zhang}}]{Gandhi2014}
{Gandhi}, P., {Lansbury}, G.~B., {Alexander}, D.~M., {et~al.} 2014, \apj, 792,
  117

\bibitem[{{Garmire} {et~al.}(2003){Garmire}, {Bautz}, {Ford}, {Nousek}, \&
  {Ricker}}]{Garmire2003}
{Garmire}, G.~P., {Bautz}, M.~W., {Ford}, P.~G., {Nousek}, J.~A., \& {Ricker},
  Jr., G.~R. 2003, in \procspie, Vol. 4851, X-Ray and Gamma-Ray Telescopes and
  Instruments for Astronomy., ed. J.~E. {Truemper} \& H.~D. {Tananbaum}, 28--44

\bibitem[{{Gehrels}(1986)}]{Gehrels1986}
{Gehrels}, N. 1986, \apj, 303, 336

\bibitem[{{George} \& {Fabian}(1991)}]{George1991}
{George}, I.~M., \& {Fabian}, A.~C. 1991, \mnras, 249, 352

\bibitem[{{Gibson} {et~al.}(2008){Gibson}, {Brandt}, \&
  {Schneider}}]{Gibson2008a}
{Gibson}, R.~R., {Brandt}, W.~N., \& {Schneider}, D.~P. 2008, \apj, 685, 773

\bibitem[{{Gibson} {et~al.}(2009){Gibson}, {Jiang}, {Brandt}, {Hall}, {Shen},
  {Wu}, {Anderson}, {Schneider}, {Vanden Berk}, {Gallagher}, {Fan}, \&
  {York}}]{Gibson2009}
{Gibson}, R.~R., {Jiang}, L., {Brandt}, W.~N., {et~al.} 2009, \apj, 692, 758

\bibitem[{{Gilfanov} \& {Merloni}(2014)}]{Gilfanov2014}
{Gilfanov}, M., \& {Merloni}, A. 2014, \ssr, 183, 121

\bibitem[{{Hao} {et~al.}(2011){Hao}, {Elvis}, {Civano}, \&
  {Lawrence}}]{Hao2011}
{Hao}, H., {Elvis}, M., {Civano}, F., \& {Lawrence}, A. 2011, \apj, 733, 108

\bibitem[{{Hewett} \& {Foltz}(2003)}]{Hewett2003}
{Hewett}, P.~C., \& {Foltz}, C.~B. 2003, \aj, 125, 1784

\bibitem[{{Hewett} {et~al.}(1995){Hewett}, {Foltz}, \& {Chaffee}}]{Hewett1995}
{Hewett}, P.~C., {Foltz}, C.~B., \& {Chaffee}, F.~H. 1995, \aj, 109, 1498

\bibitem[{{Ivezi{\'c}} {et~al.}(2002){Ivezi{\'c}}, {Menou}, {Knapp}, {Strauss},
  {Lupton}, {Vanden Berk}, {Richards}, {Tremonti}, {Weinstein}, {Anderson},
  {Bahcall}, {Becker}, {Bernardi}, {Blanton}, {Eisenstein}, {Fan},
  {Finkbeiner}, {Finlator}, {Frieman}, {Gunn}, {Hall}, {Kim}, {Kinkhabwala},
  {Narayanan}, {Rockosi}, {Schlegel}, {Schneider}, {Strateva}, {SubbaRao},
  {Thakar}, {Voges}, {White}, {Yanny}, {Brinkmann}, {Doi}, {Fukugita},
  {Hennessy}, {Munn}, {Nichol}, \& {York}}]{Ivezi2002}
{Ivezi{\'c}}, {\v Z}., {Menou}, K., {Knapp}, G.~R., {et~al.} 2002, \aj, 124,
  2364

\bibitem[{{Jiang} {et~al.}(2010){Jiang}, {Fan}, {Brandt}, {Carilli}, {Egami},
  {Hines}, {Kurk}, {Richards}, {Shen}, {Strauss}, {Vestergaard}, \&
  {Walter}}]{Jiang2010}
{Jiang}, L., {Fan}, X., {Brandt}, W.~N., {et~al.} 2010, \nat, 464, 380

\bibitem[{{Just} {et~al.}(2007){Just}, {Brandt}, {Shemmer}, {Steffen},
  {Schneider}, {Chartas}, \& {Garmire}}]{Just2007}
{Just}, D.~W., {Brandt}, W.~N., {Shemmer}, O., {et~al.} 2007, \apj, 665, 1004

\bibitem[{{Kraft} {et~al.}(1991){Kraft}, {Burrows}, \& {Nousek}}]{Kraft1991}
{Kraft}, R.~P., {Burrows}, D.~N., \& {Nousek}, J.~A. 1991, \apj, 374, 344

\bibitem[{{Leighly} {et~al.}(2007{\natexlab{a}}){Leighly}, {Halpern},
  {Jenkins}, \& {Casebeer}}]{Leighly2007a}
{Leighly}, K.~M., {Halpern}, J.~P., {Jenkins}, E.~B., \& {Casebeer}, D.
  2007{\natexlab{a}}, \apjs, 173, 1

\bibitem[{{Leighly} {et~al.}(2007{\natexlab{b}}){Leighly}, {Halpern},
  {Jenkins}, {Grupe}, {Choi}, \& {Prescott}}]{Leighly2007b}
{Leighly}, K.~M., {Halpern}, J.~P., {Jenkins}, E.~B., {et~al.}
  2007{\natexlab{b}}, \apj, 663, 103

\bibitem[{{Luo} {et~al.}(2013){Luo}, {Brandt}, {Alexander}, {Harrison},
  {Stern}, {Bauer}, {Boggs}, {Christensen}, {Comastri}, {Craig}, {Fabian},
  {Farrah}, {Fiore}, {Fuerst}, {Grefenstette}, {Hailey}, {Hickox}, {Madsen},
  {Matt}, {Ogle}, {Risaliti}, {Saez}, {Teng}, {Walton}, \& {Zhang}}]{Luo2013}
{Luo}, B., {Brandt}, W.~N., {Alexander}, D.~M., {et~al.} 2013, \apj, 772, 153

\bibitem[{{Luo} {et~al.}(2014){Luo}, {Brandt}, {Alexander}, {Stern}, {Teng},
  {Ar{\'e}valo}, {Bauer}, {Boggs}, {Christensen}, {Comastri}, {Craig},
  {Farrah}, {Gandhi}, {Hailey}, {Harrison}, {Koss}, {Ogle}, {Puccetti}, {Saez},
  {Scott}, {Walton}, \& {Zhang}}]{Luo2014}
---. 2014, \apj, 794, 70

\bibitem[{{Luo} {et~al.}(2015){Luo}, {Brandt}, {Hall}, {Wu}, {Anderson},
  {Garmire}, {Gibson}, {Plotkin}, {Richards}, {Schneider}, {Shemmer}, \&
  {Shen}}]{Luo2015}
{Luo}, B., {Brandt}, W.~N., {Hall}, P.~B., {et~al.} 2015, \apj, 805, 122

\bibitem[{{Luo} {et~al.}(2017){Luo}, {Brandt}, {Xue}, {Lehmer}, {Alexander},
  {Bauer}, {Vito}, {Yang}, {Basu-Zych}, {Comastri}, {Gilli}, {Gu},
  {Hornschemeier}, {Koekemoer}, {Liu}, {Mainieri}, {Paolillo}, {Ranalli},
  {Rosati}, {Schneider}, {Shemmer}, {Smail}, {Sun}, {Tozzi}, {Vignali}, \&
  {Wang}}]{Luo2017}
{Luo}, B., {Brandt}, W.~N., {Xue}, Y.~Q., {et~al.} 2017, \apjs, 228, 2

\bibitem[{{Lusso} {et~al.}(2010){Lusso}, {Comastri}, {Vignali}, {Zamorani},
  {Brusa}, {Gilli}, {Iwasawa}, {Salvato}, {Civano}, {Elvis}, {Merloni},
  {Bongiorno}, {Trump}, {Koekemoer}, {Schinnerer}, {Le Floc'h}, {Cappelluti},
  {Jahnke}, {Sargent}, {Silverman}, {Mainieri}, {Fiore}, {Bolzonella}, {Le
  F{\`e}vre}, {Garilli}, {Iovino}, {Kneib}, {Lamareille}, {Lilly}, {Mignoli},
  {Scodeggio}, \& {Vergani}}]{Lusso2010}
{Lusso}, E., {Comastri}, A., {Vignali}, C., {et~al.} 2010, \aap, 512, A34

\bibitem[{{Lyons}(1991)}]{Lyons1991}
{Lyons}, L. 1991, {A Practical Guide to Data Analysis for Physical Science
  Students}, 107

\bibitem[{{Lyu} {et~al.}(2017){Lyu}, {Rieke}, \& {Shi}}]{Lyu2017}
{Lyu}, J., {Rieke}, G.~H., \& {Shi}, Y. 2017, \apj, 835, 257

\bibitem[{{Martin} {et~al.}(2005){Martin}, {Fanson}, {Schiminovich},
  {Morrissey}, {Friedman}, {Barlow}, {Conrow}, {Grange}, {Jelinsky},
  {Milliard}, {Siegmund}, {Bianchi}, {Byun}, {Donas}, {Forster}, {Heckman},
  {Lee}, {Madore}, {Malina}, {Neff}, {Rich}, {Small}, {Surber}, {Szalay},
  {Welsh}, \& {Wyder}}]{Martin2005}
{Martin}, D.~C., {Fanson}, J., {Schiminovich}, D., {et~al.} 2005, \apjl, 619,
  L1

\bibitem[{{Mateos} {et~al.}(2010){Mateos}, {Carrera}, {Page}, {Watson},
  {Corral}, {Tedds}, {Ebrero}, {Krumpe}, {Schwope}, \& {Ceballos}}]{Mateos2010}
{Mateos}, S., {Carrera}, F.~J., {Page}, M.~J., {et~al.} 2010, \aap, 510, A35

\bibitem[{{Matt} {et~al.}(2012){Matt}, {Bianchi}, {Guainazzi}, {Barcons}, \&
  {Panessa}}]{Matt2012}
{Matt}, G., {Bianchi}, S., {Guainazzi}, M., {Barcons}, X., \& {Panessa}, F.
  2012, \aap, 540, A111

\bibitem[{{Matthews} {et~al.}(2016){Matthews}, {Knigge}, {Long}, {Sim},
  {Higginbottom}, \& {Mangham}}]{Matthews2016}
{Matthews}, J.~H., {Knigge}, C., {Long}, K.~S., {et~al.} 2016, \mnras, 458, 293

\bibitem[{{Miller} {et~al.}(2011){Miller}, {Brandt}, {Schneider}, {Gibson},
  {Steffen}, \& {Wu}}]{Miller2011}
{Miller}, B.~P., {Brandt}, W.~N., {Schneider}, D.~P., {et~al.} 2011, \apj, 726,
  20

\bibitem[{{Murphy} \& {Yaqoob}(2009)}]{Murphy2009}
{Murphy}, K.~D., \& {Yaqoob}, T. 2009, \mnras, 397, 1549

\bibitem[{{Murray} {et~al.}(1995){Murray}, {Chiang}, {Grossman}, \&
  {Voit}}]{Murray1995}
{Murray}, N., {Chiang}, J., {Grossman}, S.~A., \& {Voit}, G.~M. 1995, \apj,
  451, 498

\bibitem[{{O'Donnell}(1994)}]{Donnell1994}
{O'Donnell}, J.~E. 1994, \apj, 422, 158

\bibitem[{{Park} {et~al.}(2006){Park}, {Kashyap}, {Siemiginowska}, {van Dyk},
  {Zezas}, {Heinke}, \& {Wargelin}}]{Park2006}
{Park}, T., {Kashyap}, V.~L., {Siemiginowska}, A., {et~al.} 2006, \apj, 652,
  610

\bibitem[{{Planck Collaboration} {et~al.}(2016){Planck Collaboration}, {Ade},
  {Aghanim}, {Arnaud}, {Ashdown}, {Aumont}, {Baccigalupi}, {Banday},
  {Barreiro}, {Bartlett}, \& et~al.}]{Ade2016}
{Planck Collaboration}, {Ade}, P.~A.~R., {Aghanim}, N., {et~al.} 2016, \aap,
  594, A13

\bibitem[{{Plotkin} {et~al.}(2016){Plotkin}, {Gallo}, {Haardt}, {Miller},
  {Wood}, {Reines}, {Wu}, \& {Greene}}]{Plotkin2016}
{Plotkin}, R.~M., {Gallo}, E., {Haardt}, F., {et~al.} 2016, \apj, 825, 139

\bibitem[{{Proga}(2005)}]{Proga2005}
{Proga}, D. 2005, \apjl, 630, L9

\bibitem[{{Proga} {et~al.}(2000){Proga}, {Stone}, \& {Kallman}}]{Proga2000}
{Proga}, D., {Stone}, J.~M., \& {Kallman}, T.~R. 2000, \apj, 543, 686

\bibitem[{{Reeves} {et~al.}(1997){Reeves}, {Turner}, {Ohashi}, \&
  {Kii}}]{Reeves1997}
{Reeves}, J.~N., {Turner}, M.~J.~L., {Ohashi}, T., \& {Kii}, T. 1997, \mnras,
  292, 468

\bibitem[{{Reichard} {et~al.}(2003){Reichard}, {Richards}, {Schneider}, {Hall},
  {Tolea}, {Krolik}, {Tsvetanov}, {Vanden Berk}, {York}, {Knapp}, {Gunn}, \&
  {Brinkmann}}]{Reichard2003}
{Reichard}, T.~A., {Richards}, G.~T., {Schneider}, D.~P., {et~al.} 2003, \aj,
  125, 1711

\bibitem[{{Richards} {et~al.}(2006){Richards}, {Lacy}, {Storrie-Lombardi},
  {Hall}, {Gallagher}, {Hines}, {Fan}, {Papovich}, {Vanden Berk}, {Trammell},
  {Schneider}, {Vestergaard}, {York}, {Jester}, {Anderson}, {Budav{\'a}ri}, \&
  {Szalay}}]{Richards2006}
{Richards}, G.~T., {Lacy}, M., {Storrie-Lombardi}, L.~J., {et~al.} 2006, \apjs,
  166, 470

\bibitem[{{Rieke} {et~al.}(2004){Rieke}, {Young}, {Engelbracht}, {Kelly},
  {Low}, {Haller}, {Beeman}, {Gordon}, {Stansberry}, {Misselt}, {Cadien},
  {Morrison}, {Rivlis}, {Latter}, {Noriega-Crespo}, {Padgett}, {Stapelfeldt},
  {Hines}, {Egami}, {Muzerolle}, {Alonso-Herrero}, {Blaylock}, {Dole}, {Hinz},
  {Le Floc'h}, {Papovich}, {P{\'e}rez-Gonz{\'a}lez}, {Smith}, {Su}, {Bennett},
  {Frayer}, {Henderson}, {Lu}, {Masci}, {Pesenson}, {Rebull}, {Rho}, {Keene},
  {Stolovy}, {Wachter}, {Wheaton}, {Werner}, \& {Richards}}]{Rieke2004}
{Rieke}, G.~H., {Young}, E.~T., {Engelbracht}, C.~W., {et~al.} 2004, \apjs,
  154, 25

\bibitem[{{Rovilos} {et~al.}(2014){Rovilos}, {Georgantopoulos}, {Akylas},
  {Aird}, {Alexander}, {Comastri}, {Del Moro}, {Gandhi}, {Georgakakis},
  {Harrison}, \& {Mullaney}}]{Rovilos2014}
{Rovilos}, E., {Georgantopoulos}, I., {Akylas}, A., {et~al.} 2014, \mnras, 438,
  494

\bibitem[{{Schneider} {et~al.}(2010){Schneider}, {Richards}, {Hall}, {Strauss},
  {Anderson}, {Boroson}, {Ross}, {Shen}, {Brandt}, {Fan}, {Inada}, {Jester},
  {Knapp}, {Krawczyk}, {Thakar}, {Vanden Berk}, {Voges}, {Yanny}, {York},
  {Bahcall}, {Bizyaev}, {Blanton}, {Brewington}, {Brinkmann}, {Eisenstein},
  {Frieman}, {Fukugita}, {Gray}, {Gunn}, {Hibon}, {Ivezi{\'c}}, {Kent}, {Kron},
  {Lee}, {Lupton}, {Malanushenko}, {Malanushenko}, {Oravetz}, {Pan}, {Pier},
  {Price}, {Saxe}, {Schlegel}, {Simmons}, {Snedden}, {SubbaRao}, {Szalay}, \&
  {Weinberg}}]{Schneider2010}
{Schneider}, D.~P., {Richards}, G.~T., {Hall}, P.~B., {et~al.} 2010, \aj, 139,
  2360

\bibitem[{{Scott} {et~al.}(2011){Scott}, {Stewart}, {Mateos}, {Alexander},
  {Hutton}, \& {Ward}}]{Scott2011}
{Scott}, A.~E., {Stewart}, G.~C., {Mateos}, S., {et~al.} 2011, \mnras, 417, 992

\bibitem[{{Shen} {et~al.}(2011){Shen}, {Richards}, {Strauss}, {Hall},
  {Schneider}, {Snedden}, {Bizyaev}, {Brewington}, {Malanushenko},
  {Malanushenko}, {Oravetz}, {Pan}, \& {Simmons}}]{Shen2011}
{Shen}, Y., {Richards}, G.~T., {Strauss}, M.~A., {et~al.} 2011, \apjs, 194, 45

\bibitem[{{Simmonds} {et~al.}(2016){Simmonds}, {Bauer}, {Thuan}, {Izotov},
  {Stern}, \& {Harrison}}]{Simmonds2016}
{Simmonds}, C., {Bauer}, F.~E., {Thuan}, T.~X., {et~al.} 2016, \aap, 596, A64

\bibitem[{{Skrutskie} {et~al.}(2006){Skrutskie}, {Cutri}, {Stiening},
  {Weinberg}, {Schneider}, {Carpenter}, {Beichman}, {Capps}, {Chester},
  {Elias}, {Huchra}, {Liebert}, {Lonsdale}, {Monet}, {Price}, {Seitzer},
  {Jarrett}, {Kirkpatrick}, {Gizis}, {Howard}, {Evans}, {Fowler}, {Fullmer},
  {Hurt}, {Light}, {Kopan}, {Marsh}, {McCallon}, {Tam}, {Van Dyk}, \&
  {Wheelock}}]{Skrutskie2006}
{Skrutskie}, M.~F., {Cutri}, R.~M., {Stiening}, R., {et~al.} 2006, \aj, 131,
  1163

\bibitem[{{Sprayberry} \& {Foltz}(1992)}]{Sprayberry1992}
{Sprayberry}, D., \& {Foltz}, C.~B. 1992, \apj, 390, 39

\bibitem[{{Steffen} {et~al.}(2006){Steffen}, {Strateva}, {Brandt}, {Alexander},
  {Koekemoer}, {Lehmer}, {Schneider}, \& {Vignali}}]{Steffen2006}
{Steffen}, A.~T., {Strateva}, I., {Brandt}, W.~N., {et~al.} 2006, \aj, 131,
  2826

\bibitem[{{Stocke} {et~al.}(1992){Stocke}, {Morris}, {Weymann}, \&
  {Foltz}}]{Stocke1992}
{Stocke}, J.~T., {Morris}, S.~L., {Weymann}, R.~J., \& {Foltz}, C.~B. 1992,
  \apj, 396, 487

\bibitem[{{Teng} {et~al.}(2014){Teng}, {Brandt}, {Harrison}, {Luo},
  {Alexander}, {Bauer}, {Boggs}, {Christensen}, {Comastri}, {Craig}, {Fabian},
  {Farrah}, {Fiore}, {Gandhi}, {Grefenstette}, {Hailey}, {Hickox}, {Madsen},
  {Ptak}, {Rigby}, {Risaliti}, {Saez}, {Stern}, {Veilleux}, {Walton}, {Wik}, \&
  {Zhang}}]{Teng2014}
{Teng}, S.~H., {Brandt}, W.~N., {Harrison}, F.~A., {et~al.} 2014, \apj, 785, 19

\bibitem[{{Trump} {et~al.}(2006){Trump}, {Hall}, {Reichard}, {Richards},
  {Schneider}, {Vanden Berk}, {Knapp}, {Anderson}, {Fan}, {Brinkman},
  {Kleinman}, \& {Nitta}}]{Trump2006}
{Trump}, J.~R., {Hall}, P.~B., {Reichard}, T.~A., {et~al.} 2006, \apjs, 165, 1

\bibitem[{{Vanden Berk} {et~al.}(2001){Vanden Berk}, {Richards}, {Bauer},
  {Strauss}, {Schneider}, {Heckman}, {York}, {Hall}, {Fan}, {Knapp},
  {Anderson}, {Annis}, {Bahcall}, {Bernardi}, {Briggs}, {Brinkmann}, {Brunner},
  {Burles}, {Carey}, {Castander}, {Connolly}, {Crocker}, {Csabai}, {Doi},
  {Finkbeiner}, {Friedman}, {Frieman}, {Fukugita}, {Gunn}, {Hennessy},
  {Ivezi{\'c}}, {Kent}, {Kunszt}, {Lamb}, {Leger}, {Long}, {Loveday}, {Lupton},
  {Meiksin}, {Merelli}, {Munn}, {Newberg}, {Newcomb}, {Nichol}, {Owen}, {Pier},
  {Pope}, {Rockosi}, {Schlegel}, {Siegmund}, {Smee}, {Snir}, {Stoughton},
  {Stubbs}, {SubbaRao}, {Szalay}, {Szokoly}, {Tremonti}, {Uomoto}, {Waddell},
  {Yanny}, \& {Zheng}}]{Vanden2001}
{Vanden Berk}, D.~E., {Richards}, G.~T., {Bauer}, A., {et~al.} 2001, \aj, 122,
  549

\bibitem[{{Weymann} {et~al.}(1991){Weymann}, {Morris}, {Foltz}, \&
  {Hewett}}]{Weymann1991}
{Weymann}, R.~J., {Morris}, S.~L., {Foltz}, C.~B., \& {Hewett}, P.~C. 1991,
  \apj, 373, 23

\bibitem[{{Wright} {et~al.}(2010){Wright}, {Eisenhardt}, {Mainzer}, {Ressler},
  {Cutri}, {Jarrett}, {Kirkpatrick}, {Padgett}, {McMillan}, {Skrutskie},
  {Stanford}, {Cohen}, {Walker}, {Mather}, {Leisawitz}, {Gautier}, {McLean},
  {Benford}, {Lonsdale}, {Blain}, {Mendez}, {Irace}, {Duval}, {Liu}, {Royer},
  {Heinrichsen}, {Howard}, {Shannon}, {Kendall}, {Walsh}, {Larsen}, {Cardon},
  {Schick}, {Schwalm}, {Abid}, {Fabinsky}, {Naes}, \& {Tsai}}]{Wright2010}
{Wright}, E.~L., {Eisenhardt}, P.~R.~M., {Mainzer}, A.~K., {et~al.} 2010, \aj,
  140, 1868

\bibitem[{{Xue} {et~al.}(2011){Xue}, {Luo}, {Brandt}, {Bauer}, {Lehmer},
  {Broos}, {Schneider}, {Alexander}, {Brusa}, {Comastri}, {Fabian}, {Gilli},
  {Hasinger}, {Hornschemeier}, {Koekemoer}, {Liu}, {Mainieri}, {Paolillo},
  {Rafferty}, {Rosati}, {Shemmer}, {Silverman}, {Smail}, {Tozzi}, \&
  {Vignali}}]{Xue2011}
{Xue}, Y.~Q., {Luo}, B., {Brandt}, W.~N., {et~al.} 2011, \apjs, 195, 10

\bibitem[{{York} {et~al.}(2000){York}, {Adelman}, {Anderson}, {Anderson},
  {Annis}, {Bahcall}, {Bakken}, {Barkhouser}, {Bastian}, {Berman}, {Boroski},
  {Bracker}, {Briegel}, {Briggs}, {Brinkmann}, {Brunner}, {Burles}, {Carey},
  {Carr}, {Castander}, {Chen}, {Colestock}, {Connolly}, {Crocker}, {Csabai},
  {Czarapata}, {Davis}, {Doi}, {Dombeck}, {Eisenstein}, {Ellman}, {Elms},
  {Evans}, {Fan}, {Federwitz}, {Fiscelli}, {Friedman}, {Frieman}, {Fukugita},
  {Gillespie}, {Gunn}, {Gurbani}, {de Haas}, {Haldeman}, {Harris}, {Hayes},
  {Heckman}, {Hennessy}, {Hindsley}, {Holm}, {Holmgren}, {Huang}, {Hull},
  {Husby}, {Ichikawa}, {Ichikawa}, {Ivezi{\'c}}, {Kent}, {Kim}, {Kinney},
  {Klaene}, {Kleinman}, {Kleinman}, {Knapp}, {Korienek}, {Kron}, {Kunszt},
  {Lamb}, {Lee}, {Leger}, {Limmongkol}, {Lindenmeyer}, {Long}, {Loomis},
  {Loveday}, {Lucinio}, {Lupton}, {MacKinnon}, {Mannery}, {Mantsch}, {Margon},
  {McGehee}, {McKay}, {Meiksin}, {Merelli}, {Monet}, {Munn}, {Narayanan},
  {Nash}, {Neilsen}, {Neswold}, {Newberg}, {Nichol}, {Nicinski}, {Nonino},
  {Okada}, {Okamura}, {Ostriker}, {Owen}, {Pauls}, {Peoples}, {Peterson},
  {Petravick}, {Pier}, {Pope}, {Pordes}, {Prosapio}, {Rechenmacher}, {Quinn},
  {Richards}, {Richmond}, {Rivetta}, {Rockosi}, {Ruthmansdorfer}, {Sandford},
  {Schlegel}, {Schneider}, {Sekiguchi}, {Sergey}, {Shimasaku}, {Siegmund},
  {Smee}, {Smith}, {Snedden}, {Stone}, {Stoughton}, {Strauss}, {Stubbs},
  {SubbaRao}, {Szalay}, {Szapudi}, {Szokoly}, {Thakar}, {Tremonti}, {Tucker},
  {Uomoto}, {Vanden Berk}, {Vogeley}, {Waddell}, {Wang}, {Watanabe},
  {Weinberg}, {Yanny}, {Yasuda}, \& {SDSS Collaboration}}]{York2000}
{York}, D.~G., {Adelman}, J., {Anderson}, Jr., J.~E., {et~al.} 2000, \aj, 120,
  1579

\bibitem[{{Young} {et~al.}(2009){Young}, {Elvis}, \& {Risaliti}}]{Young2009}
{Young}, M., {Elvis}, M., \& {Risaliti}, G. 2009, \apjs, 183, 17

\bibitem[{{Yuan} \& {Wills}(2003)}]{Yuan2003}
{Yuan}, M.~J., \& {Wills}, B.~J. 2003, \apjl, 593, L11

\end{thebibliography}

\clearpage
\begin{deluxetable}{lcccccccc}
\tabletypesize{\scriptsize}
\tablewidth{0pt}
\tablecaption{X-ray Photometric Properties 
}
\tablehead{
\colhead{Object Name}   &
\multicolumn{2}{c}{Net Counts}   &
\colhead{Band}   &
\colhead{$\Gamma_{\rm eff}$}   &
\colhead{$\Gamma_{\rm XSPEC}$}   &
\multicolumn{2}{c}{Flux ($10^{-14}$\flux)}  &
\colhead{$\log L_{\rm X}$ (\lum)}    \\
\cline{2-3}   \cline{7-8} \\
\colhead{(LBQS B)} &
\colhead{0.5--2 keV} &
\colhead{2--8 keV} &
\colhead{Ratio} &
\colhead{} &
\colhead{} &
\colhead{0.5--2 keV} &
\colhead{2--8 keV} &
\colhead{2--10 keV}        \\
\colhead{(1)}          &
\colhead{(2)}          &
\colhead{(3)}          &
\colhead{(4)}          &
\colhead{(5)}          &
\colhead{(6)}          &
\colhead{(7)}          &
\colhead{(8)}          &
\colhead{(9)}      
}
\startdata
$ 0021-0213$&$   6.4_{-2.6}^{+4.6}$&$   5.9_{-2.7}^{+4.9}$&$     0.91_{-0.44}^{+1.57}$&$   1.0_{-0.7}^{+0.5}$&$   1.2_{-0.6}^{+0.4}$&$   0.12$&$   0.46$&$   43.9$\\
$ 1203+1530$&$   7.2_{-2.7}^{+3.9}$&$   1.9_{-1.4}^{+2.9}$&$     0.26_{-0.11}^{+0.68}$&$   2.2_{-0.9}^{+0.9}$&$   2.0_{-0.7}^{+0.6}$&$   0.21$&$   0.17$&$   43.6$\\
$ 1212+1445$&$   4.1_{-2.0}^{+3.3}$&$   5.2_{-2.4}^{+3.7}$&$     1.28_{-0.58}^{+2.35}$&$   0.7_{-0.8}^{+0.6}$&$   0.5_{-0.6}^{+0.6}$&$   0.13$&$   0.81$&$   43.8$\\
$ 1235+1453$&$   4.0_{-2.0}^{+3.3}$&$   6.6_{-2.7}^{+4.0}$&$     1.63_{-0.71}^{+2.73}$&$   0.4_{-0.8}^{+0.5}$&$   0.8_{-0.6}^{+0.6}$&$   0.11$&$   0.96$&$   44.1$\\
$ 1442-0011$&$   5.1_{-2.2}^{+3.5}$&$   2.0_{-1.4}^{+2.9}$&$     0.39_{-0.16}^{+1.00}$&$   1.9_{-0.8}^{+0.9}$&$   1.8_{-0.7}^{+0.7}$&$   0.19$&$   0.24$&$   44.0$\\
$ 1443+0141$&$  24.9_{-5.1}^{+6.2}$&$  10.7_{-3.4}^{+4.7}$&$     0.43_{-0.30}^{+0.65}$&$   1.7_{-0.3}^{+0.4}$&$   1.2_{-0.3}^{+0.3}$&$   0.79$&$   1.25$&$   44.7$\\
$ 2201-1834$&$                <4.0$&$                <4.1$&$                      ...$&$                 ...$&$                 ...$&$  <0.13$&$  <0.70$&$  <43.8$\\

\enddata
\tablecomments{
Col. (1): object name.
Cols. (2)--(3): net counts in the observed soft (0.5--2 keV) and hard (2--8 keV) bands.
Col. (4): ratio between the \hbox{soft-band} and \hbox{hard-band} counts, 
the symbol "..." indicates that the source is undetected in both the soft and hard bands. 
Col. (5): effective power-law photon index derived from the photometric approach.
These values were used in our analyses and discussions.
Col. (6): best-fit power-law photon index obtained from spectral fitting. 
Cols. (7)--(8): Galactic absorption-corrected flux in the soft (0.5--2~keV) and hard (2--8 keV) bands. 
Col. (9): logarithm of the rest-frame 2--10 keV luminosity, derived from the 0.5--2~keV flux.
}
\label{tbl-xbasic}
\end{deluxetable}

\begin{deluxetable}{lcccccccccc}
\tabletypesize{\scriptsize}

\tablewidth{0pt}
\tablecaption{X-ray and Optical Properties}
\tablehead{
\colhead{Object Name}                  &
\colhead{Count Rate}                   &
\colhead{$f_{\rm 2keV,soft}$}                   &
\colhead{$f_{\rm 2keV,hard}$}                   &
\colhead{$f_{\rm 2500~\textup{\AA}}$}                   &
\colhead{$\log L_{\rm 2500~\textup{\AA}}$}              &
\colhead{$\alpha_{\rm OX,soft}$}                   &
\colhead{$\alpha_{\rm OX,hard}$}                   &
\colhead{$\Delta\alpha_{\rm OX,soft}(\sigma)$}                   &
\colhead{$\Delta\alpha_{\rm OX,hard}(\sigma)$}                   &
\colhead{$f_{\rm weak}$}                  \\
\colhead{(LBQS B)}   &
\colhead{(0.5--2~keV)}   &
\colhead{}   &
\colhead{}   &
\colhead{}   &
\colhead{}   &
\colhead{}   &
\colhead{}   &
\colhead{}   &
\colhead{}   &
\colhead{}   \\
\colhead{(1)}         &
\colhead{(2)}         &
\colhead{(3)}         &
\colhead{(4)}         &
\colhead{(5)}        &
\colhead{(6)}         &
\colhead{(7)}         &
\colhead{(8)}         &
\colhead{(9)}         &
\colhead{(10)}        &
\colhead{(11)}   
}
\startdata
$ 0021-0213$&$     0.18_{-0.07}^{+0.13}$&$    0.32$&$    1.81$&$    2.34$&$   31.46$&$    -2.25_{-0.07}^{+0.12}$&$    -1.96_{-0.08}^{+0.14}$&$ -0.58(3.95)$&$ -0.29(1.99)$&$        5.7_{-3.2}^{+3.3}$\\
$ 1203+1530$&$     0.38_{-0.14}^{+0.21}$&$    0.90$&$    0.71$&$    1.68$&$   31.05$&$    -2.02_{-0.06}^{+0.09}$&$    -2.06_{-0.12}^{+0.25}$&$ -0.41(2.79)$&$ -0.45(3.07)$&$     14.7_{-11.5}^{+16.4}$\\
$ 1212+1445$&$     0.26_{-0.13}^{+0.21}$&$    0.31$&$    2.28$&$    4.46$&$   31.47$&$    -2.36_{-0.08}^{+0.13}$&$    -2.03_{-0.08}^{+0.12}$&$ -0.69(4.73)$&$ -0.36(2.44)$&$        8.5_{-4.4}^{+4.9}$\\
$ 1235+1453$&$     0.22_{-0.11}^{+0.18}$&$    0.19$&$    3.54$&$    1.48$&$   31.38$&$    -2.26_{-0.08}^{+0.14}$&$    -1.77_{-0.07}^{+0.10}$&$ -0.60(4.13)$&$ -0.11(0.77)$&$        2.0_{-0.9}^{+1.0}$\\
$ 1442-0011$&$     0.34_{-0.15}^{+0.24}$&$    0.86$&$    1.13$&$    3.45$&$   31.60$&$    -2.15_{-0.07}^{+0.11}$&$    -2.11_{-0.12}^{+0.24}$&$ -0.46(3.15)$&$ -0.41(2.83)$&$      11.9_{-9.2}^{+12.6}$\\
$ 1443+0141$&$     1.55_{-0.32}^{+0.39}$&$    3.50$&$    6.12$&$    2.28$&$   31.50$&$    -1.85_{-0.03}^{+0.04}$&$    -1.75_{-0.05}^{+0.07}$&$ -0.17(1.16)$&$ -0.08(0.53)$&$        1.6_{-0.6}^{+0.6}$\\
$ 2201-1834$&$                    <0.29$&$   <0.37$&$   <2.29$&$    9.40$&$   31.88$&$                   <-2.46$&$                   <-2.15$&$<-0.73(4.99)$&$<-0.42(2.91)$&$                    >12.8$\\
\enddata
\tablecomments{
Col. (1): object name.
Col. (2): 0.5--2~keV count rate in units of 10$^{-3}$~s$^{-1}$.
Col. (3): \hbox{soft-band} derived flux density at rest-frame 2~keV in units of $10^{-32}$~\mflux, derived using the measured $\Gamma_{\rm eff}$ and the \hbox{soft-band} flux.
Col. (4): \hbox{hard-band} derived flux density at rest-frame 2~keV in units of $10^{-32}$~\mflux, derived using $\Gamma=2$ and the \hbox{hard-band} flux.
Col. (5): flux density at rest-frame 2500~\AA\ in units of $10^{-27}$~\mflux, adopted from \cite{Gallagher2006}.
Col. (6): logarithm of the rest-frame 2500~\AA\ monochromatic luminosity in units of \mlum, adopted from \cite{Gallagher2006}.
Col. (7): \hbox{soft-band} derived $\alpha_{\rm OX}$ parameter, derived using $f_{\rm2keV,soft}$.
Col. (8): \hbox{hard-band} derived $\alpha_{\rm OX}$ parameter, derived using $f_{\rm2keV,hard}$.
Col. (9): difference between the \hbox{soft-band} derived $\alpha_{\rm OX}$ and the
expected $\alpha_{\rm OX}$ from the \citet{Steffen2006}
\hbox{$\alpha_{\rm OX}$--$L_{\rm 2500~{\textup{\AA}}}$} relation. 
The statistical significance of this difference,
measured in units of the $\alpha_{\rm OX}$ rms scatter in Table~5 of 
\citet{Steffen2006}, is given in the parenthesis.
Col. (10): difference between the hard band derived $\alpha_{\rm OX}$ and the
expected $\alpha_{\rm OX}$.
Col. (11): factor of \hbox{hard-band} \xray\ weakness in accordance with $\Delta\alpha_{\rm OX,hard}$.
The errors were derived from the errors of $\alpha_{\rm OX,hard}$.
}
\label{tbl-aox}
\end{deluxetable}

\end{document}